\documentclass[aps,twocolumn,pra,footinbib,floatfix,showpacs,
showkeys,superscriptaddress]{revtex4-1}
\usepackage{epsf,color,colordvi} 
\usepackage{graphicx}
\usepackage[fulladjust]{marginnote}
\usepackage{ulem}
\usepackage{amsmath}
\usepackage{times}
\usepackage{subfigure}
\usepackage{float}
\usepackage{placeins}
\usepackage{comment}

\begin{document} 

\title{Additive decomposition of iterative quantum search operations in the Grover-type algorithm}
%\title{Additive decomposition scheme of an iterative quantum search algorithm 
%of the Grover-type and its application to reducing the search load}

\author{F.M. Toyama}
\email{toyama@cc.kyoto-su.ac.jp}
\affiliation{Department of Computer Science,  Kyoto Sangyo University, Kyoto 603-8555, Japan} 
\author{W. van Dijk}
\email{vandijk@physics.mcmaster.ca}
\affiliation{Department of Physics, Redeemer University College, Ancaster, Ontario L0K 1J4, Canada and \\
Department of Physics and Astronomy, McMaster University, Hamilton, Ontario, Canada L8S 4M1 }
%\author{W. van Dijk}
%\email{vandijk@physics.mcmaster.ca}
%\affiliation{Physics Department, Redeemer University College,
% Ancaster, Ontario L9K 1J4, Canada}
%\affiliation{Department of Physics and Astronomy, McMaster University,
% Hamilton, Ontario L8S 4M1, Canada}

\date{\today}

\begin{abstract}
In the Grover-type quantum search process a search operator is iteratively applied, say, $k$ times, 
on the initial database state. 
We present an additive decomposition scheme such that the iteration process is 
expressed, in the computational space, as a linear combination of $k$ operators, each of which consists of a single Grover-search followed by an overall phase-rotation.
The value of $k$ and the rotation phase are the same as those determined in the framework of the search with certainty.
We further show that the final state can be expressed in terms of a single oracle operator
of the Grover-search and  phase-rotation factors.
We discuss how the additive form can be utilized  so that it effectively reduces
the computational load  of the iterative search, and we propose an effective shortcut gate that realizes
the same outcome as the iterative search.  
\end{abstract}  

\date{\today}



\maketitle

\section{Introduction}
\label{sec1} 
In the Grover-type quantum search process~\cite{grover96, grover97,grover97a,grover98, grover05, grover06} a search operator  
is iteratively applied on the initial database state that consists of $N=2^n$ unordered basis states to search for $M$ target 
states, $n$ being the number of qubit-registers. The phase matching 
method~\cite{toyama13, long99, li01, biham99, li07} for the Grover quantum search 
algorithm has been extensively studied and shown to be effective in improving the success
probability $P_k(\lambda)$, where $\lambda=M/N$ is the ratio of the number of target states to the number of database states and $k$ is 
the number of iterations.  In spite of the impressive efficacy of this method for most values of $\lambda$, it is less so when $\lambda\ll 1$.

In Ref.~\cite{toyama13}, we investigated the problem of an exact search with 
the success probability $P_k(\lambda)=1$ for any value of $\lambda$, on the basis of
the phase-matched search-operator $G_N(\alpha) \equiv W_N(-\alpha)U_N(\alpha)$, where $U_N(\alpha)$ 
is the oracle operator, $W_N(-\alpha)$ is the diffusion operator and $\alpha$ is the matching phase.
The search operator used in the original Grover search is a special case of the above with $\alpha=\pi$. 
We assumed that  $\lambda$ is known preliminarily. Then the phase 
matching method enabled us to accomplish the exact search.
We gave analytic forms of optimal number of searches 
$k$ and the matching phase $\alpha_k$ for the exact search for the entire range of $0< \lambda \le 1$.    
We showed that $k = k_G$ or $k = k_G + 1$, where $k_G$ 
is the optimal number of searches of the original Grover algorithm (see Fig.~1 of Ref.~\cite{toyama13}).  
Recall that, in the original version of the Grover search, $P_{k_G}(\lambda)=1$ can be satisfied only for 
$\lambda=1/4$ and $\lambda=1$~\cite{diao10}.

The purpose of this paper is to propose a way of further expediting the search process by effectively 
reducing the search load.  
We first derive, in the $N$-dimensional computational space, an additive decomposition 
scheme for the $k$-iterative search state $|\phi_k\rangle = G_N^k(\alpha) |\phi_0 \rangle$, where $|\phi_0 \rangle$
and $|\phi_k \rangle$ are respectively the initial database state and the final state.
The decomposed form includes a linear combination of $k$ components each of which consists of a single 
Grover-search operator $G_N(\alpha)$ followed by an overall phase-rotation.  
In this scheme, the number of oracle operations can be reduced to unity.
The phase-rotation parameter is determined by $\alpha$ and $\lambda$,
or by the optimal number of iterations $k$  for the exact search. The phase-rotation parameter
can be determined preliminarily. 
In the decomposed form, $|\phi_k\rangle$ can be expressed as a simple superposition of 
$|\phi_0\rangle$ and $|\phi_1\rangle $ with superposition coefficients that are determined by the 
phase-rotation parameter.
This enables us to obtain $|\phi_k\rangle$  with only the information of $|\phi_1 \rangle$.
Furthermore, it is noteworthy that $|\phi_1 \rangle$ can be expressed in terms of only the oracle operation 
$U_N(-\alpha) |\phi_0\rangle$ and the phase-rotation parameter, without the diffusion operator $W_N(\alpha)$. 
This enables us to obtain $|\phi_k \rangle$ in terms of only $U_N(-\alpha)$ and the phase-rotation 
parameter. This is a key feature of the additive decomposition scheme.

As we show in due course, however, the reduced operator itself obtained in the decomposition 
is, unlike $G^{k}_N(\alpha)$, not unitary except in the case of $n=1\ (N=2)$, although the norm of 
$|\phi_k \rangle$ is preserved in the decomposition.
By making use of the reduced form for $|\phi_k\rangle$ and on the basis of the search with certain outcome
algorithm~\cite{toyama13},
however, we can define a unitary gate that directly transforms $|\phi_0\rangle$ to $|\phi_k \rangle$.
The unitary transformation provides a shortcut from $|\phi_0\rangle$ to $|\phi_k\rangle$, 
by bypassing the Grover-type iterative searches. The unitary transformation so obtained is much simpler than 
the corresponding $k$-iterative search
operator $G_N^k(\alpha)$. 
We will show this by examining the matrix representations of the shortcut operator and the iterative search operator
$G^k_N(\alpha)$.

In Sec. II, we first give a brief review of the exact search algorithm that was developed in 
Ref.~\cite{toyama13} and subsequently derive 
the decomposition scheme.  In Sec. III, we present a shortcut scheme for 
the iterative search in the framework of the exact search.  Section IV contains a summary. 
In Appendix A we give an illustration of the decomposition scheme, 
in Appendix B we illustrate the unitary shortcut operator and in Appendix C we discuss a parallel processing
scheme implied by the reduced operator based on the decomposition.

\section{Additive decomposition scheme of an iterative search algorithm}
\label{sec2}
%======================Subsection A====================================
\subsection{Iterative-search with $P_k(\lambda)=1$ for $0<\lambda \leq 1$}
\label{subsec:IIA}
We first give a brief review of the iterative search algorithm that yields exactly 
$P(\lambda)=1$ for any $0< \lambda \ (=M/N) \leq 1$~\cite{toyama13}.
In the computational space of $N=2^n$ dimensions, 
a modified Grover algorithm based on the phase matching method is represented by the oracle operator $U_{N}(\alpha)$ 
and the diffusion operator $W_{N}\left(-\alpha\right)$,
%-------------eqs.(1)(2)-------------------------------------------------------------------------------------------
\begin{eqnarray} 
  U_{N}(\alpha)&=&I_N-(1-e^{i\alpha})\sum_{l=0}^{M-1}|t_l\rangle\langle t_l|,  \label{eq:1} \\
  W_{N}\left(-\alpha\right)&=& H^{\otimes n}  \left[ I_N e^{-i\alpha}+(1-e^{-i\alpha}) 
      | 0^{\otimes n}\rangle \langle 0^{\otimes n} | \right]   H^{\otimes n}    \nonumber  \\
  &=&I_N e^{-i\alpha}+(1-e^{-i\alpha})|\phi_0\rangle\langle \phi_0 |,  \label{eq:2}
\end{eqnarray}
%--------------------------------------------------------------------------------------------------------------------
where $\alpha$ is the matching phase, suffix $N$ indicates that the operators are of the computational space,  
$|0^{\otimes n}\rangle$ is the $n$-qubit initialized register state  
and $H$ is the Walsh-Hadamard transformation. 
The unstructured initial database state $|\phi_0\rangle$ is defined by 
$|\phi_0\rangle = H^{\otimes n} |0^{\otimes n}\rangle=(1/\sqrt{N}) \sum_{l=0}^{N-1}|\omega_l\rangle$,
where $|\omega_{l}\rangle$ are the computational basis states. 
The $|\phi_0\rangle$ can also be written as  
$|\phi_{0}\rangle =  \sqrt{1-\lambda} |R\rangle + \sqrt{\lambda} |T\rangle$,
where $|T\rangle$ is the uniform superposition of target states $|t_l\rangle$, i.e., 
$|T\rangle=\frac{1}{\sqrt M} \sum_{l=0}^{M-1}|t_l\rangle$, and  $|R\rangle$ is that of the remaining states  
$|r_l\rangle$, i.e., $|R\rangle = \frac{1}{\sqrt {N-M}}\sum_{l=0}^{N-M-1}|r_l\rangle$.  \

A $k$ time search is done by $k$ iterative operations of the kernel operator 
$G_{N}(\alpha) \equiv W_{N}(-\alpha) U_{N}(\alpha)$ on $|\phi_0\rangle$, i.e.,
%---------------------eq.(3)---------------------------------------
\begin{equation}   \label{eq:3}
  |\phi_k \rangle = G_{N}^k(\alpha) |\phi_0 \rangle,  \\
\end{equation}
%--------------------------------------------------------------------
where $|\phi_k \rangle$ is the state obtained by the $k$ iterative search. 
In Ref.~\cite{toyama13}, we showed that the matching phase $\alpha$ for the exact search with 
$P_k(\lambda)=1$ can be determined in terms of $\lambda$ and $k$ as
%---------------------eq.(4)---------------------------------------------------------------------------
\begin{equation} \label{eq:4}
  \alpha_k(\lambda)  = \arccos\left[ 1- \frac{1-\cos(\pi/(2k+1))}{\lambda} \right].   \\
\end{equation}
%-------------------------------------------------------------------------------------------------------
For a given value of $\lambda$, we first determine the optimal number $k$ of the iterations  
as the smallest integer that is compatible with 
%---------------------eq.(5)-----------------------------------------------------
\begin{equation} \label{eq:5}
  k \geq \frac{\pi-\arccos(1-2\lambda)}{2\arccos(1-2\lambda)}.  \\
\end{equation}
%----------------------------------------------------------------------------------
The optimal $k$ is a staircase function of $\lambda$ (see Fig.~1 of Ref.~\cite{toyama13}). 
%%---------------------eq.(6)------------------------------------------------------------------------------------------------------
%\begin{equation} \label{eq:6}
%  \lambda \geq \lambda_{\mathrm{min}} \equiv \frac{1}{2} \left[1-\cos\left(\frac{\pi}{2k+1}\right)\right].  \\
%\end{equation}
%%------------------------------------------------------------------------------------------------------------------------------------
If we know $\lambda$ preliminarily, by using Eqs.~(\ref{eq:4}) and (\ref{eq:5}) we can determine the 
optimal $k$ and $\alpha_k(\lambda)$ for the search with certainty. 
As mentioned in the introduction, the optimal $k$ for the exact search is always equal to, or greater by one 
than, that of the original Grover search depending on the value of $\lambda$.

%==================Subsection B================================
\subsection{Novel relation between $G_N(\alpha)$ and $G_N^{\dag}(\alpha)$}
\label{subsec:IIB}
We present a useful relationship between $G_N(\alpha)$ and $G_N^{\dag}(\alpha)$ 
for the purpose of applying it to the decomposition scheme for the iterative search mentioned in the 
previous sub-section.  
By using Eqs.~(\ref{eq:1}) and (\ref{eq:2}), it can be verified that the initial state $|\phi_0 \rangle$ 
is an eigenstate of $G_{N}(\alpha) + G^{\dag}_{N}(\alpha)$ with the eigenvalue 
$\epsilon = 2\left[1-\lambda(1-\cos\alpha) \right] $, namely
%---------------------eq.(6)----------------------------------------------------------------------------------------------------------------
\begin{eqnarray} \label{eq:6}
  \left[ G_{N}(\alpha) + G^{\dag}_{N}(\alpha) \right] |\phi_0 \rangle &=& \epsilon \ \ |\phi_0 \rangle \nonumber \\
  &=& 2 \left[1-\lambda(1-\cos\alpha) \right] 
|\phi_0 \rangle. 
\end{eqnarray}
%----------------------------------------------------------------------------------------------------------------------------------------------
In fact, $|T\rangle$ and  $|R\rangle$ are individually eigenstates of 
$G_{N}(\alpha) + G^{\dag}_{N}(\alpha)$ belonging to the same eigenvalue $\epsilon$. 
Hence, $|\phi_0 \rangle$ is an eigenstate of $G_{N}(\alpha) + G^{\dag}_{N}(\alpha)$ belonging to
 the same eigenvalue $\epsilon$.

First, we show that  $\epsilon$ is equal to the trace of the search operator 
$G_{[2]}(\alpha) \equiv W_{[2]}(-\alpha)U_{[2]}(\alpha)$ represented in the two-dimensional 
space spanned by the basis $\{ |R\rangle, |T\rangle \} $.
In this space the search operator $G_{[2]}(\alpha)$ is represented 
as~\cite{toyama13}, 
%---------------------eq.(7)-----------------------------------------------------------------------------------------------------------------------
\begin{eqnarray}\label{eq:7}
&G_{[2]}&(\alpha)   \nonumber \\
&=&\!\!\!
\begin{pmatrix}
      1-(1-e^{-i\alpha})\lambda & -(1-e^{i\alpha})\sqrt{\lambda(1-\lambda)} \\
      (1-e^{-i\alpha})\sqrt{\lambda(1-\lambda)} &  1-(1-e^{i\alpha})\lambda 
       \end{pmatrix}.
\end{eqnarray}
%------------------------------------------------------------------------------------------------------------------------------------------------------
It turns out that  $\mathrm{Tr}  G_{[2]}(\alpha)  = 2 \left[1-\lambda(1 - \cos\alpha) \right] 
= \epsilon_{+} + \epsilon_{-} = \epsilon$, where $\epsilon_{\pm}$ are the two eigenvalues of $G_{[2]}(\alpha)$.
The $\epsilon_{\pm}$ are given by
%--------------------eq.(8)------------------------------------------------------------------------------------------------------------
\begin{eqnarray}\label{eq:8} 
  &\epsilon_\pm &=1 - \lambda \left(1 - \cos\alpha \right)   \nonumber  \\
  & & \pm i \sqrt{   \lambda \left(1- \cos\alpha  \right)  \left[ 2 - \lambda \left( 1 - \cos\alpha \right)  \right]   } .    
\end{eqnarray} 
%----------------------------------------------------------------------------------------------------------------------------------------
Since $G_{[2]}(\alpha)$ is unitary, of course $|\epsilon_\pm|=1$. 
The two eigenvalues $\epsilon_{+}$ and $\epsilon_{-}$ are complex conjugate of each other and satisfy 
the relation $\epsilon_{+} \epsilon_{-}=1$.
We can express $\epsilon_\pm$ as $\epsilon_\pm = e^{\pm i \theta} $, where  
%--------------------eq.(9)------------------------------------------------------------------------------------
 \begin{equation} \label{eq:9}
  \theta = \arctan\left( \frac{\sqrt{x(2-x)}}{1-x} \right), \ \ \ x\equiv \lambda (1-\cos\alpha). \\
\end{equation} 
%------------------------------------------------------------------------------------------------------------------
The phase $\theta$ is determined by $\alpha$ and $\lambda$. From Eq.~(\ref{eq:8}), 
it is seen that $\theta$ is simply related to $\alpha$ and $\lambda$ as,
%--------------------eq.(10)-----------------
 \begin{equation} \label{eq:10}
 \cos\theta=1-\lambda(1-\cos\alpha). \\
\end{equation} 
%-------------------------------------------------

Equation~(\ref{eq:6}) can be rewritten as
%--------------------eq.(11)------------------------------------------------------------------------------------------------------------
\begin{eqnarray} 
&\left[ G_{N}(\alpha) + G^{\dag}_{N}(\alpha) \right] |\phi_0 \rangle&   \nonumber \\
  &=& \!\!\!\!\!\!\!\!\!\!\!\!\!\!\!\!\!\!\!\!\!\!\!\!\! 
  \left[\mathrm{Tr} G_{[2]}(\alpha) \right] |\phi_0 \rangle  = (\epsilon_{+} + \epsilon_{-} ) |\phi_0 \rangle  \label{eq:11} \\
  &=& \!\!\!\!\!\!\!\!\!\!\!\!\!\!\!\!\!\!\!\!\!\!\!\!\! (e^{i \theta} +e^{-i \theta}) |\phi_0 \rangle. \nonumber 
\end{eqnarray} 
%-----------------------------------------------------------------------------------------------------------------------------------------
It should be stressed that Eq.~(\ref{eq:11}) holds as an eigenvalue equation.  On the other hand,
in the two-dimensional representation of Eq.~(\ref{eq:7}), the relation 
$G_{[2]}(\alpha) + G^{\dag}_{[2]}(\alpha)= [\mathrm{Tr} G_{[2]}(\alpha)] I_2 =(\epsilon_{+} + \epsilon_{-})I_2 $ 
holds as a relation between the two-dimensional unitary matrices $G_{[2]}(\alpha)$ and $G^{\dag}_{[2]}(\alpha)$. 
This relation can be proved by using the two dimensional Cayley-Hamilton theorem for 
unitary $G_{[2]}(\alpha)$ with its $\det G_{[2]}(\alpha)=1$. 

Next, we extend the relation~(\ref{eq:11})  to an arbitrary number of searches $k$.
By operating $G_N(\alpha)$ + $G_N^{\dag}(\alpha)$ on both sides of  Eq.~(\ref{eq:11}) from the left, 
we obtain,
%--------------------eq.(12)-------------------------------------------------------------------------------------------------------
\begin{eqnarray}
&\left\{ G_{N}(\alpha)^2 + \left[G_N^{\dag}(\alpha)\right]^2 + 2 I_N \right\}& |\phi_0 \rangle  \nonumber \\
&=& \!\!\!\!\!\!\!\!\!\!\!\!\!\!\!\!\!\!\!\!\!\!\!\!\!\!\!\!\!\!\!\!\!\!\!\!
\mathrm{Tr}G_{[2]}(\alpha)  \left[ G_{N}(\alpha) + G_N^{\dag}(\alpha) \right]  |\phi_0 \rangle  \nonumber \\  
&=& \!\!\!\!\!\!\!\!\!\!\!\!\!\!\!\!\!\!\!\!\!\!\!\!\!\!\!\!\!\!\!\!\!\!\!\!
\left[ \mathrm{Tr}G_{[2]}(\alpha)\right]^2  |\phi_0 \rangle,        \label{eq:12}   
\end{eqnarray} 
%-------------------------------------------------------------------------------------------------------------------------------------
where the unitarity of $G_{N}(\alpha)$ was used.  
Equation~(\ref{eq:12})  can be rewritten as
%--------------------eq.(13)-------------------------------------------------------------------------------------------------------------------------
\begin{eqnarray} \label{eq:13} 
&&\left\{ G_{N}^2(\alpha) + \left[G_N^{\dag}(\alpha)\right]^2 \right\} \! |\phi_0 \rangle    
= \left\{ \left[ \mathrm{Tr}G_{[2]}(\alpha)\right]^2  - 2 \right\} \! |\phi_0 \rangle   \nonumber \\
&&\ \ \ \ \ \ \ \ \ \ \ \ \ \ \ \ \ \ \ \ \ \ \ = \left[ (\epsilon_{+} +\epsilon_{-})^2  - 2\right] |\phi_0 \rangle     \nonumber 
= \left( \epsilon^2_{+} +\epsilon^2_{-} \right) |\phi_0 \rangle        \nonumber \\
&&\ \ \ \ \ \ \ \ \ \ \ \ \ \ \ \ \ \ \ \ \ \ \ = \left[ \mathrm{Tr}G_{[2]}^2(\alpha) \right] |\phi_0 \rangle,  
\end{eqnarray} 
%------------------------------------------------------------------------------------------------------------------------------------------------------------
where $\epsilon_{+} \epsilon_{-} =1$ was used. Since $[ G^{\dag}_{N}(\alpha) ]^2=[ G^2_{N}(\alpha)]^{\dag}$ we obtain,
%---------------------eq.(14)-----------------------------------------------------------------------------------------------------------------------------
\begin{eqnarray} \label{eq:14}
& &\!\!\!\!\!\!\!\!\!\!\!\!\!\!\!\!\!\!\!\!\!\!\!\!\!\!\!  \left\{ G_{N}^2(\alpha) + \left[ G^2_{N}(\alpha) \right]^{\dag} \right\} |\phi_0 \rangle  \nonumber \\   
  &=& \left[\mathrm{Tr}G^2_{[2]}(\alpha)\right] |\phi_0 \rangle = \left(\epsilon^2_{+}  + \epsilon^2_{-}\right) |\phi_0 \rangle.  
\end{eqnarray} 
%------------------------------------------------------------------------------------------------------------------------------------------------------------
By repeating the same procedure, we obtain
%---------------------eq.(15)--------------------------------------------------------------------------------------------------
\begin{eqnarray} \label{eq:15} 
\left\{ G_{N}^k(\alpha) + \left[ G^k_{N}(\alpha) \right]^{\dag}  \right\} |\phi_0 \rangle    
= \left[\mathrm{Tr} G_{[2]}^{k}(\alpha) \right] |\phi_0 \rangle \nonumber \\   
 = \left(\epsilon^k_{+}  + \epsilon^k_{-}\right) |\phi_0 \rangle = \left(e^{i k \theta}  + e^{- i k \theta} \right) |\phi_0 \rangle  
\end{eqnarray}
%-------------------------------------------------------------------------------------------------------------------------------------------------
for any number  $k$ of searches. 
For $k=0$, Eq.~(\ref{eq:15}) should be understood as $( I_{N} + I^{\dag}_{N} ) 
|\phi_0 \rangle  =  \left(Tr I_{2}\right) |\phi_0 \rangle = 2|\phi_0 \rangle$.
The decomposition scheme presented in this subsection holds for any search specified by $k$, $N$ 
and $\lambda$, as far as $\theta$ is determined by the relation (\ref{eq:10}). 
The search does not always have to be the exact search of Ref.~\cite{toyama13} reviewed 
in subsection ~\ref{subsec:IIA}. 

For the exact search of Ref.~\cite{toyama13}, $\alpha_k(\lambda)$ is given by Eq.~(\ref{eq:4}).
In this case, from Eqs.~(\ref{eq:4}) and  (\ref{eq:10}) it turns out that the phase $\theta_k$  is 
determined by the optimal $k$ alone as \cite{tdn13}
%-------------------eq.(16)---------------------
\begin{equation} \label{eq:16}
   \theta_k = \frac{\pi}{2k + 1}. \\
\end{equation} 
%---------------------------------------------------
Since the optimal $k$ is a staircase function of $\lambda$, $\theta_k$ is also so.

%==================== Subsection C =================================================
\subsection{ Additive decomposition scheme of $G_N^k(\alpha)|\phi_0\rangle$} 
\label{subsec:IIC}
We now construct the decomposition scheme on the basis of Eq.~(\ref{eq:15}). 
By applying $G_N^k(\alpha)$ to the left of both sides of Eq.~(\ref{eq:15}), we obtain,
%---------------------eq.(17)------------------------------------------------------------------------------------------------------
\begin{equation} \label{eq:17} 
 \left[ G_{N}^{2k}(\alpha) + I_N \right] |\phi_0 \rangle  =  \left(e^{i k \theta}  + e^{- i k \theta} \right)  G_N^k(\alpha) 
|\phi_0 \rangle.      
\end{equation}
%-------------------------------------------------------------------------------------------------------------------------------------
Equation~(\ref{eq:17}) can be rewritten as
%---------------------eq.(18)----------------------------------------------------------------------------------------------------------------------------
\begin{eqnarray} \label{eq:18}  
|\phi_{2k} \rangle &\equiv & G_{N}^{2k}(\alpha)  |\phi_0 \rangle  \nonumber  \\
& = & \left[ \left(e^{i k \theta}  + e^{- i k \theta} \right) G_N^k(\alpha) + e^{i \pi} I_N  \right] |\phi_0 \rangle   \\
%&&= \left[ e^{i k \theta} G_N^k(\alpha) + e^{-i k \theta}  G_N^k(\alpha)  + e^{i \pi} I_N \right] |\phi_0 \rangle   \nonumber \\
&\equiv &  e^{i k \theta} |\phi_{k} \rangle +  e^{-i k \theta} |\phi_{k} \rangle + e^{i \pi} |\phi_0 \rangle.
\end{eqnarray}
%------------------------------------------------------------------------------------------------------------------------------------------------------------
Equation~(\ref{eq:18}) implies that the $2k$ search can be decomposed into a sum of two $k$-iterative 
searches,  $e^{ i k \theta} G_N^k(\alpha) |\phi_0 \rangle$ and $e^{- i k \theta} G_N^k(\alpha) |\phi_0 \rangle$, 
and a constant phase transformation $ e^{i \pi} I_N |\phi_0 \rangle$. 
The first two operations consist of $k$ searches of $G_N^k(\alpha)$ followed by the unitary overall phase rotation 
$e^{\pm i k \theta} I_N=e^{\pm i k \theta} I_2^{\otimes n}$. 
With the aid of Eq.~(\ref{eq:15}), it can directly be confirmed that the norm of the decomposed form  
$e^{i k \theta} |\phi_{k} \rangle +  e^{-i k \theta} |\phi_{k} \rangle + e^{i \pi} |\phi_0 \rangle$ is unity, where the norm 
of $|\phi_{k} \rangle$ is of course unity.
Equation~(\ref{eq:18}) gives a basic decomposition scheme for an even number of iterations.

By operating $G_N(\alpha)$ on Eq.~(\ref{eq:18}) from the left, 
a basic decomposition scheme for an odd number of iterations is obtained  as
%-----------------------eq.(19)-------------------------------------------------------------------------------------------------------------------------
\begin{eqnarray}  \label{eq:19} 
&&\!\!\!\!\!\!\!\!\!\!\!\!\!\!\!\!\!\!\! |\phi_{2k+1}\rangle \equiv G_{N}^{2k+1}(\alpha) |\phi_0 \rangle    \nonumber \\  
&&\ = G_{N}(\alpha) \left(e^{i k \theta} |\phi_{k} \rangle +  e^{-i k \theta} |\phi_{k} \rangle 
+ e^{i \pi} |\phi_0 \rangle \right)     \nonumber \\  
&&\  \equiv     e^{i k\theta} |\phi_{k+1} \rangle + e^{-i k\theta} |\phi_{k+1}\rangle    
 + e^{i \pi} |\phi_{1}\rangle. 
\end{eqnarray}
%---------------------------------------------------------------------------------------------------------------------------------------------------------- 
Similarly to the case of an even number of iterations of Eq.~(\ref{eq:18}), it can directly be
 confirmed that the norm of the decomposed 
expression $e^{i k\theta} |\phi_{k+1} \rangle + e^{-i k\theta} |\phi_{k+1}\rangle + e^{i \pi} |\phi_{1}\rangle$ is unity, where 
the norms of $|\phi_{k+1} \rangle$ and $|\phi_{1} \rangle$ are both unity.    

By using the two basic decomposition schemes of Eqs.~(\ref{eq:18}) and  (\ref{eq:19}), we can decompose 
a search of any number of iterations into a linear combination\sout{s} of a single Grover search followed by overall 
phase rotations of the form of $e^{\pm i m \theta} I_N (m: \rm{an\ \ integer})$ and a constant phase rotation $e^{i \pi} I_N$.
We give below explicit forms of the decompositions for $k=1,\dots,6$,
%-------------------------eqs.(20)(21)(22)(23)(24)(25)--------------------------------------------------------------------------------------------
\begin{eqnarray} 
 |\phi_1\rangle &=& G_{N}(\alpha)|\phi_0 \rangle   \nonumber  \\
                       &=& \left[ \left( e^{i \theta} + e^{-i \theta} \right) I_N + e^{i\pi}G^{\dag}_N(\alpha) \right]   |\phi_0 \rangle,     \label{eq:20}  \\
\!\!\!\!\!\!\!\! |\phi_2\rangle &=& G^{2}_{N}(\alpha) |\phi_0 \rangle   \nonumber \\
                       &=& \left[ \left( e^{i \theta} + e^{-i \theta} \right) G_N(\alpha) + e^{i\pi} I_N \right] |\phi_0 \rangle,   \label{eq:21} \\ 
\!\!\!\!\!\!\!\! |\phi_3\rangle &=& G^{3}_{N}(\alpha) |\phi_0 \rangle   \nonumber \\
                       &=& \left[ \left( e^{2i\theta} + e^{-2i\theta} + 1 \right) G_N(\alpha)   \right.    \nonumber  \\
                       & & \left. \ \ \ \ \ \ \ + \left(e^{i\theta} + e^{-i\theta}\right)  e^{i\pi} I_N \right] |\phi_0 \rangle,   \label{eq:22}\\ 
\!\!\!\!\!\!\!\! |\phi_4\rangle &=& G^{4}_{N}(\alpha) |\phi_0 \rangle   \nonumber \\
                       &=& \left[ \left( e^{3i\theta} + e^{-3i\theta} + e^{i\theta} + e^{-i\theta} \right) G_N(\alpha)   \right.  \nonumber  \\
                  & & \left. \ \ \  + \left(e^{2i\theta} + e^{-2i\theta} +1 \right)  e^{i\pi} I_N \right] |\phi_0 \rangle,      \label{eq:23}\\
\!\!\!\!\!\!\!\! |\phi_5\rangle &=& G^{5}_{N}(\alpha) |\phi_0 \rangle   \nonumber \\
                       &=& \left[ \left( e^{4i\theta} + e^{-4i\theta} + e^{2i\theta} + e^{-2i\theta} + 1 \right) G_N(\alpha)   \right.  \nonumber  \\    
                & &\!\!\!\!\!  \left. + \left(e^{3i\theta} + e^{-3i\theta} + e^{i\theta} + e^{-i\theta} \right)  e^{i\pi} I_N \right] |\phi_0 \rangle,    \label{eq:24} \\ 
%\text{and}   \nonumber \\          
\!\!\!\!\!\!\!\! |\phi_6\rangle &=& G^{6}_{N}(\alpha) |\phi_0 \rangle   \nonumber \\
                       &=& \left[ \left( e^{5i\theta} + e^{-5i\theta} + e^{3i\theta} + e^{-3i\theta} + e^{i\theta} + e^{-i\theta} \right) G_N(\alpha)   \right.  \nonumber  \\    
                & &\!\!\!\!\!  \left. + \left(e^{4i\theta} + e^{-4i\theta} + e^{2i\theta} + e^{-2i\theta} +1\right)  e^{i\pi} I_N \right] |\phi_0 \rangle \label{eq:25}.                                                                
\end{eqnarray}
%------------------------------------------------------------------------------------------------------------------------------------------------------------ 
For $k\geq 2$, it is seen, from Eqs.~(\ref{eq:20}) to (\ref{eq:25}), that the additive decomposition of 
$|\phi_k \rangle$ can be summarized as, 
%---------------------------eq.(26)--------------------------------------------------------------------------------------------------------
\begin{eqnarray}  \label{eq:26}
\!\!\!\! |\phi_k\rangle &=& G^{k}_{N}(\alpha) |\phi_0 \rangle   
                       = \left[ f_k(\theta) G_N(\alpha)  + f_{k-1}(\theta)e^{i\pi} I_N \right] |\phi_0 \rangle  \nonumber  \\
                       &\equiv& G^{I,k}_N(\theta, \alpha)  |\phi_0 \rangle,                                                         
\end{eqnarray}
%------------------------------------------------------------------------------------------------------------------------------------------------ 
where $G^{I,k}_N(\theta, \alpha)  \equiv f_k(\theta) G_N(\alpha)  + f_{k-1}(\theta)e^{i\pi} I_N$,
and $f_k(\theta)$ is given as,
%----------------------------eq.(27)--------------------------------------------------------------------------------------
\begin{eqnarray}  \label{eq:27}
\!\!\!\!\!\!\!\! f_k(\theta) &=& f_{k-1}(\theta) (e^{i\theta} + e^{-i\theta}) - f_{k-2}(\theta)  \  \ (k \ge 2).                                                             
\end{eqnarray}
%-------------------------------------------------------------------------------------------------------------------------------- 
It is understood that $f_0(\theta)=0$ and $f_1(\theta)=1$.             
Note that $f_k^{\ast}(\theta)=f_k(\theta)$.
The norm of the reduced form of $G^{I,k}_N(\theta, \alpha)  |\phi_0 \rangle$  of
Eq.~(\ref{eq:26})  again can be verified to be unity for any $k$ with the aid of the relation
$[G_{N}(\alpha) + G^{\dag}_{N}(\alpha)] |\phi_0\rangle=(e^{i\theta} + e^{-i\theta}) |\phi_0\rangle$
of Eq.~(\ref{eq:11}) and the identity $f^2_{k-1}(\theta) - f_k(\theta)f_{k-2}(\theta) = 1$ which follows from Eq.~(\ref{eq:27}).  
Only when $N=2$ is $G^{I,k}_N(\theta, \alpha)$ unitary.

As seen in  Eq.~(\ref{eq:20}),  $|\phi_1\rangle = G_N(\alpha) |\phi_0\rangle$ can also be expressed in terms 
of $G^{\dag}_N(\alpha)$ instead of $G_N(\alpha)$.  Equation~(\ref{eq:20}) can further be rewritten as
%------------------------------eq.(28)--------------------------------------------------------------------------------------------------
\begin{eqnarray}  
 |\phi_1\rangle &=& G_{N}(\alpha)|\phi_0 \rangle  
 = \left[ \left(e^{i \theta} + e^{-i \theta}\right) I_N + e^{i\pi}G^{\dag}_N(\alpha) \right] |\phi_0 \rangle   \nonumber  \\
%      &=& \left( e^{i \theta} + e^{-i \theta} + e^{i\pi}U_N(-\alpha) \right)   |\phi_0 \rangle,   \\  \label{eq:21f} 
      &=& \left( f_2(\theta) I_N + e^{i\pi}U_N(-\alpha) \right)   |\phi_0 \rangle,       \label{eq:28} 
\end{eqnarray}
%-------------------------------------------------------------------------------------------------------------------------------------------
where the relation $G^{\dag}_{N}(\alpha) |\phi_0 \rangle = 
U^{\dag}_{N}(\alpha)W^{\dag}_{N}(-\alpha) |\phi_0 \rangle = U_{N}(-\alpha)W_{N}(\alpha) |\phi_0 \rangle 
= U_{N}(-\alpha) |\phi_0 \rangle $ was used. Here, note that $W_{N}(\alpha) |\phi_0 \rangle = |\phi_0 \rangle$. 
Thus, in order to obtain $| \phi_1 \rangle$ we need only the oracle operation $U_{N}(-\alpha) |\phi_0 \rangle$. 
By using $|\phi_1\rangle$ of Eq.~(\ref{eq:28}) in Eq.~(\ref{eq:26}), 
we can express $| \phi_k \rangle$ as
%------------------------------eq.(29)--------------------------------------------------------------------------------------------------
\begin{eqnarray} 
&&\!\!\!\!\!\!\!\!\!\!\!\!\!\!\!\!\!\!\! |\phi_k\rangle = \left[ \left( f_k(\theta) f_2(\theta)  + e^{i \pi} f_{k-1}(\theta) \right) I_N  
 \right. \nonumber \\
&&       \ \ \ \ \ \ \ \ \ \ \ \ \ \ \ \ \ \ \ \ \  \left.   + \ e^{i \pi} f_k(\theta) U_N(-\alpha) \right] |\phi_0 \rangle   \label{eq:29}  \\ 
&&\!\!\!\!\!\!\!= \left[ g_k(\theta) I_N + h_k(\theta) U_N(-\alpha) \right] |\phi_0 \rangle,   \nonumber  \\
&&\!\!\!\!\!\!\! \equiv G_N^{I\!I,k} (\theta, \alpha) |\phi_0 \rangle,   \nonumber
\end{eqnarray}
%-------------------------------------------------------------------------------------------------------------------------------------------
where {$G_N^{I\!I,k} (\theta, \alpha) \equiv  g_k(\theta) I_N + h_k(\theta) U_N(-\alpha)$,
 $g_k(\theta) \equiv f_k(\theta) f_2(\theta) + e^{i \pi} f_{k-1}(\theta)$ and $h_k(\theta) \equiv e^{i \pi} f_k(\theta)$.
Thus $|\phi_k\rangle$ can be obtained by the oracle operation $U_N(-\alpha) | \phi_0 \rangle$ alone. 
No diffusion operation is needed in obtaining $|\phi_k \rangle$.
This is a key feature of the additive decomposition of the iterative search process.
As the diffusion operator is eliminated from $G^{I,k}_N(\theta, \alpha)$ of Eq.~(\ref{eq:26}),
 $G^{I\!I,k}_N(\theta, \alpha)$ is not unitary, even for $N=2$.

The identity for $|\phi_k\rangle$ of Eq.~(\ref{eq:29}) (or  Eq.~(\ref{eq:26})) 
holds for any search specified by $k$, $N$ and $\lambda$, as long as $\theta$ is determined by 
the relation (\ref{eq:10}).
The reduced form of Eq.~(\ref{eq:29}) is especially useful in obtaining the vector-component 
representation of $|\phi_k\rangle$. For example, when the targets are the first $M$ 
states $|\omega_0\rangle$, $\ldots$, $|\omega_{M-1}\rangle$, by using Eq.~(\ref{eq:29}) we can write 
$|\phi_k\rangle$ at first sight as,
%------------------------------eq.(30)--------------------------------------------------------------------------------------------------
\begin{equation} 
|\phi_k\rangle = (\overbrace{v_t, \ldots, v_t}^M, \overbrace{v_{nt}, \ldots, v_{nt}}^{N-M} )^T,   \label{eq:30}     
\end{equation}
%-------------------------------------------------------------------------------------------------------------------------------------------
where $T$ indicates {\it transposed} and $v_t$ and $v_{nt}$ are respectively the amplitude of the target states 
and the non-target states.  By using Eq.~(\ref{eq:29}), the amplitudes $v_t$ and $v_{nt}$ are immediately 
obtained as,
%------------------------------eqs.(31)(32)--------------------------------------------------------------------------------------------------
\begin{eqnarray} 
&&\!\!\!\!\!\!\!\!v_t = \frac{1}{\sqrt{N}} \left[ g_k(\theta)  +  h_k(\theta) e^{- i\alpha} \right],   \label{eq:31}  \\
&&\!\!\!\!\!\!\!\!v_{nt} = \frac{1}{\sqrt{N}} \left[ g_k(\theta)  +  h_k(\theta)\right].   \label{eq:32}  
\end{eqnarray}
%-------------------------------------------------------------------------------------------------------------------------------------------
In Ref.~\cite{toyama13},  we presented compact forms of the success and failure amplitudes, i.e., $d_k$ and $u_k$ respectively,
%-----------------------------eq.(33)-----------------------------------------------------------------------------------------------------------------
\begin{equation}\label{eq:33} 
\begin{split}
d_k & = \frac{\sqrt{\lambda}}{\sin(\theta/2)} 
\left\{\sin\left[ \left( k+\frac{1}{2} \right) \theta\right] - \frac{(1+e^{-i\alpha})\sin(k\theta)}{2\cos(\theta/2) } \right\}  \\ 
u_k & = \frac{ \sqrt{1-\lambda} }{\cos(\theta/2)} \cos \left[ \left( k+\frac{1}{2} \right) \theta \right].
\end{split}
\end{equation} 
%------------------------------------------------------------------------------------------------------------------------------------------------------
These amplitudes are defined in the two-dimensional space spanned by $\left\{ |R\rangle, |T\rangle \right\}$.  
It can be verified that the relations $\sqrt{M} v_t = d_k$
and $\sqrt{N-M}  v_{nt} = u_k$ hold.
 
In the exact search with $P_k(\lambda)=1$ for a given $\lambda$, the optimal $k$, 
$\alpha_k(\lambda)$ and $\theta_k$ are respectively determined by Eqs.~(\ref{eq:5}), (\ref{eq:4}) and 
(\ref{eq:16}) (or (\ref{eq:10})). 
With these values of $k$, $\alpha_k$ and $\theta_k$, $M |v_t|^2=1$ and  $v_{nt}=0$.

Before ending this section, we have to stress the following point.  Equation~(\ref{eq:26}) (or Eq.~(\ref{eq:29}))
is  an identity for the final state $|\phi_k\rangle$.  
In Appendix A, we illustrate this situation for the case of $n=3$ $(N=8)$ and the target $|t_0\rangle=|000\rangle$. 
In this case, the optimal number of searches is $k=2$ since $\lambda=1/8$.
From Eqs.~(\ref{eq:A4}) and (\ref{eq:A7}), it will be seen that Eq.~(\ref{eq:26}) is a correct identity
for $|\phi_2\rangle$, where the phase parameter $\theta_2$ contains the 
information on the two iterations.

On the other hand, the following point is especially crucial.
Although the reduced operator $G^{I,k}_N(\theta, \alpha)$ of Eq.~(\ref{eq:26}) yields $|\phi_k\rangle$, 
$G^{I,k}_N(\theta, \alpha)$ itself is not unitary except in the trivial case of $n=1 \ (N=2)$. 
Therefore $G^{I,k}_N(\theta, \alpha)$  can not in general be interpreted as a quantum mechanical 
evolution-operator.
The reason is as follows. In order to validate the unitarity of $G^{I,k}_N(\theta, \alpha)$, we need 
the relation $G_{N}(\alpha) + G^{\dag}_{N}(\alpha) = (e^{i \theta} + e^{-i \theta})I_N $ for the two unitary 
operators  $G_{N}(\alpha)$ and  $G^{\dag}_{N}(\alpha)$.  
This relation holds only for the $N=2\ (n=1)$ case. 
For $N > 2$, however, it does not hold. On the other hand, the eigenvalue equation 
$[G_{N}(\alpha) + G^{\dag}_{N}(\alpha)] |\phi_0\rangle=(e^{i\theta} + e^{-i\theta}) |\phi_0\rangle$ 
is valid for any $N$, which guarantees that the norm of $G^{I,k}_N(\theta, \alpha) |\phi_0 \rangle$ is unity. 
This is why Eq.~(\ref{eq:26}) is an identity for $|\phi_k\rangle$.

\vspace{0.2in}
%==== Section III ===================================================================
\section{Construction of a unitary transformation suggested by the additive decomposition}
%Effective search with a shortcut operator
%with the additive decomposition of $G^k_N(\alpha)$}
\label{sec3}

%==================== Subsection A ===============================================

In this section we discuss a possible scheme suggested by the additive decomposition scheme and
construct a unitary transformation on the basis of the reduced form of Eq.~(\ref{eq:29}).
As mentioned at the end of sub-section~\ref{subsec:IIC}, although the $k$ time search operator 
$G^{k}_N(\alpha)$ can be realized as a quantum-mechanical evolution-operator,
the reduced operator $G_N^{I\!I,k}(\theta,\alpha) = \left[ g_k(\theta) I_N + h_k(\theta) U_N(-\alpha) \right]$ 
of Eq.~(\ref{eq:29}) cannot be interpreted as a quantum mechanical evolution in general, even though 
$G_N^{I\!I,k}{\theta,\alpha)} |\phi_0\rangle$ is identical to $G^{k}_N(\alpha) |\phi_0\rangle$. 
The advantage of  $G^{I\!I,k}_N(\theta,\alpha)$ is that it enables us to calculate the 
$|\phi_k\rangle$ with only the single oracle operation $U_N(-\alpha)|\phi_0\rangle$, 
where the information on the $k$ time search is provided by $\theta$. No diffusion operation is needed. 
This is a remarkable point.
In what follows, we discuss how we can take advantage of  $G^{I\!I,k}_N(\theta,\alpha)$
to construct a unitary operator  that functions as a shortcut search.

According to Eq.~(\ref{eq:29}), the final search state $|\phi_k\rangle = G_N^k(\alpha)|\phi_0\rangle$ is given 
as a superposition of the initial state $|\phi_0\rangle$ and the state obtained by one-oracle operation 
$|\phi_U\rangle \equiv U_N(-\alpha) |\phi_0\rangle$.  The superposition coefficients $g_k(\theta)$ and $h_k(\theta)$
are basically sums of phase rotations  $e^{\pm i m \theta}$ and $e^{i \pi }$. 
Figure~1 shows an additive processing scheme that is suggested by the reduced form of 
Eq.~(\ref{eq:29}). It is understood that the superposition coefficients $g_k(\theta)$ and $h_k(\theta)$ are 
incorporated in the additive transformation $C_N$.

%------------------------------------------------ fig1 -----------------------------------------------------------------------------------------
\begin{widetext}
\begin{figure}[ht]
\begin{minipage}{2.\linewidth}
  \centering
   \resizebox{5.in}{!}{\includegraphics{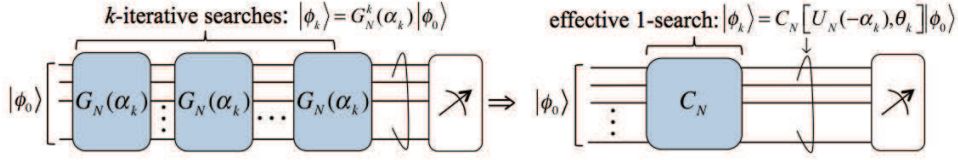}}
   \caption{ (Color online) 
    An additive processing scheme suggested by the reduced form of Eq.~(\ref{eq:29}), where the final search state 
    $|\phi_k\rangle$ is represented as a superposition of $|\phi_0\rangle$ and $|\phi_U\rangle \equiv U_N(-\alpha)|\phi_0\rangle$
    with the coefficients $g_k(\theta)$ and $h_k(\theta)$.  The unitary transformation 
    $C_N[U_N(-\alpha_k), \theta_k]$ incorporates the superposition coefficients.
     }
 \label{fig:1}
 \end{minipage}
\end{figure}
\end{widetext}
%---------------------------------------------------------------------------------------------------------------------------------------------

We now consider a certain search with $\lambda$ given in advance so that $k$, $\alpha_k$, and $\theta_k$ are determined from Eqs.~(\ref{eq:5}), (\ref{eq:4}), and (\ref{eq:16}), respectively.  The additive processing scheme shown in Fig.~\ref{fig:1} is then written as,
%-----------------------------eq.(34)-------------------------------------------------------------------------------------------------------------
\begin{eqnarray}  \label{eq:34}
|\phi_k\rangle =  C_N[U_N(-\alpha_k), \theta_k]   |\phi_0 \rangle,
\end{eqnarray}
%------------------------------------------------------------------------------------------------------------------------------------------------------     
where the unitary transformation $C_N$ can be written as,
%------------------------------eq.(35)------------------------------------------------------------------------------------------------------------
\begin{eqnarray}  \label{eq:35}
&&\!\!\!\!\!\!\!\!\! C_N[U_N(-\alpha_k), \theta_k]   \nonumber  \\
&=& |\phi^{(0)}_k \rangle \langle \phi^{(0)}_0| + |\phi^{(1)}_k \rangle \langle \phi^{(1)}_0 | 
 + \cdots + |\phi^{(N-1)}_k \rangle \langle \phi^{(N-1)}_0 |    \nonumber \\  
 &=& \left[ g_k(\theta_k)  I_N  
+ h_k(\theta_k) U_N(-\alpha_k) \right] |\phi_0 \rangle \langle \phi_0|     \nonumber \\
&&  \ \ \ \ \ \ \ \ \ + |\phi^{(1)}_k \rangle \langle \phi^{(1)}_0 |   + \cdots + |\phi^{(N-1)}_k \rangle \langle \phi^{(N-1)}_0 |.    
%\!\!\!\!\!\!\!\!\!\!\!&\equiv& |\phi^{(0)}_k \rangle \langle \phi^{(0)}_0| + |\phi^{(1)}_k \rangle \langle \phi^{(1)}_0 | 
%+ \cdots + |\phi^{(N-1)}_k \rangle \langle \phi^{(N-1)}_0 |,    \nonumber         
\end{eqnarray}
%-------------------------------------------------------------------------------------------------------------------------------------------------------
The  $C_N$ can be interpreted as a shortcut operator. 
In essence, the first term $|\phi^{(0)}_k \rangle \langle \phi^{(0)}_0| \equiv |\phi_k \rangle \langle \phi_0|$ acts as 
the transformation from $|\phi_0\rangle$ to $|\phi_k\rangle$. Other terms 
$|\phi^{(i)}_k \rangle \langle \phi^{(i)}_0 | \ \ (i=1,\dots,N-1)$ are needed to complete the unitarity of $C_N$\cite{messiah61}.

Let us now consider the significance of  $C_N$ as a search operator.
The reduction in the search speed is in particular important  for searches when $\lambda$ is very small.
To make the point clearer, as a simple example, we consider the case of a single target 
$|t_0\rangle=|\omega_0\rangle=|00 \cdots 0 \rangle$. 
In this case, from Eq.~(\ref{eq:29}) $|\phi_k \rangle = [g_k(\theta_k)I_N + h_k(\theta_k) U_N(-\alpha_k)] |\phi_0\rangle 
= v_t |\omega_0\rangle$. 
Hence, the computational basis vectors $|\omega_i\rangle \ \ (i=1,\dots,N-1)$ can be chosen as 
the mutually orthonormal states $|\phi^{(i)}_k \rangle \ \ (i=1,\dots, N-1)$.
On the other hand, the mutually orthonormal states $|\phi^{(i)}_0 \rangle \ \ (i=1,\dots, N-1)$ about
$|\phi^{(0)}_0 \rangle \equiv |\phi_0 \rangle$ can be obtained by means of the Gram-Schmidt orthogonalization
with the subsidiarily chosen $(N-1)$ linearly independent  vectors. Thus, using Eq.~(\ref{eq:29}), we can write 
$C_N$ of Eq.~(\ref{eq:34}) as,
%------------------------------eq.(36)--------------------------------------------------------------------------------------------------
\begin{eqnarray}       \label{eq:36}
&&\!\!\!\!\!C_N\!\left[U_N(-\alpha_k), \theta_k \right]   \nonumber  \\
&& \ \ \ \ \   = \left[ g_k(\theta_k)  I_N  
+ h_k(\theta_k) U_N(-\alpha_k) \right] |\phi_0 \rangle \langle \phi_0|   \nonumber  \\
&& \ \ \ \ \ \   + |\omega_1 \rangle \langle \phi^{(1)}_0 |    
+ \cdots + |\omega_{N-1} \rangle \langle \phi^{(N-1)}_0 |.                                       
\end{eqnarray}
%--------------------------------------------------------------------------------------------------------------------------------------------  
The first term in Eq.~(\ref{eq:36}) acts as a search operator for the target $|\omega_0\rangle$.  
Other terms $|\omega_i \rangle \langle \phi^{(i)}_0 |$ are also expressed in terms of 
$ |\omega_i \rangle  \langle \omega_j |$ ($i=1,\dots,N-1, j=0,\dots, N-1$). 
In this sense, we can regard $C_N$ as a shortcut search operator that consists of $U_N(-\alpha_k)$.

In order to analyze the complexity of $C_N$ it would be meaningful to examine its matrix representation 
in the computational basis. 
In the present case, the vector representations of the orthonormal set $|\phi^{(i)}_k\rangle \ \ (i=0,\dots,N-1)$ 
are given as 
%------------------------------eq.(37)--------------------------------------------------------------------------------------------------
\begin{equation}   \label{eq:37}
\begin{split}
|\phi^{(0)}_k \rangle &=|\phi_k \rangle=v_t |\omega_0\rangle = (v_t, 0, \ldots, 0, 0, 0)^T \\
|\phi^{(1)}_k \rangle & =|\omega_1\rangle=(0, 1, 0, \ldots, 0)^T \\
|\phi^{(2)}_k \rangle  & = |\omega_2\rangle =(0, 0, 1, \ldots, 0)^T  \\
\vdots \\
| \phi^{(N-1)}_k \rangle & = |\omega_{N-1}\rangle = (0, 0, \ldots, 0, 0,1)^T.
\end{split}
\end{equation}
%------------------------------------------------------------------------------------------------------------------------------------------
On the other hand, $$|\phi^{(0)}_0\rangle \equiv |\phi_0\rangle=(1/\sqrt{N}) (1, \cdots, 1,1,1)^T. \nonumber$$ 
Therefore, 
%------------------------------eq.(38)--------------------------------------------------------------------------------------------------
\begin{equation}   \label{eq:38}
\begin{split}
 |\omega_0\rangle & =(1, 0, \ldots, 0,0,0)^T \\
 |\omega_1\rangle & =(0, 1, \cdots, 0,0,0)^T \\
 \vdots \\
|\omega_{N-2}\rangle & =(0, 0, \cdots, 0,1,0)^T
\end{split}
\end{equation}
%-----------------------------------------------------------------------------------------------------------------------------------------
can be taken as the subsidiarily chosen linearly independent states.

In Appendix B, we give an explicit illustration how the matrix representation of $C_N$ can be constructed
along the line stated above.  As an example, we illustrate the case of $n=3$ $(N=8)$, $\lambda=1/8$, $k=2$ 
and the target  $|t_0\rangle=|\omega_0\rangle=|000\rangle$, which is the case considered in Appendix A.
As seen in Eq.~(\ref{eq:B10}) and Eq.~(\ref{eq:B11}), the matrix representation of the shortcut search operator 
$C_8[U_8(-\alpha_2), \theta_2]$ is much simpler than that of the iterative search operator $G^2_8(\alpha_2)$.
The elements of the first row of $C_8[U_8(-\alpha_2), \theta_2]$ is the representation of the first term 
$|\phi^{(0)}_2 \rangle \langle \phi^{(0)}_0 |=v_t |\omega_0\rangle \langle \phi_0|$ of Eq.~(\ref{eq:B1}).
All elements other than those of the first row depend on the choice of the orthonormal states $|\phi^{(i)}_0\rangle \ \
(i=1,\ldots,N-1)$. Only the first row of $G^2_8(\alpha_2)$ is the same as that of $C_8[U_8(-\alpha_2), \theta_2]$. 
On the other hand, the sum of the elements of each row other than the first row is zero in both $G^2_8(\alpha_2)$ 
and $C_8[U_8(-\alpha_2), \theta_2]$.  
Hence, $G^2_8(\alpha_2) |\phi_0\rangle = 
C_8[U_8(-\alpha_2), \theta_2] |\phi_0\rangle = |\phi_2 \rangle=v_t |\omega_0\rangle$ is guaranteed. 
In conclusion, it seems that the shortcut search operator $C_N$ results in significant reduction of the computational load, 
compared with that of the $k$-iterative search operator $G_N^k(\alpha_k)$.  \

In general, in constructing $C_N[U_N(-\alpha_k), \theta_k]$, we only need the information on the single oracle 
operation $U_N(-\alpha_k) |\phi_0\rangle$. The framer of the quantum search knows the oracle operator $U_N$ 
of iterative searches.  Therefore,  they can also construct the shortcut search operator 
$C_N[U_N(-\alpha_k), \theta_k]$.  Let us now consider what kind of quantum search can be done. 
The quantum search that we are proposing here can be used as follows. Suppose that the framer implements the 
quantum circuit corresponding to the shortcut unitary transformation $C_N[U_N(-\alpha_k), \theta_k]$ 
and asks someone else (a third one) to find a marked state by giving the information of $\lambda$.  
Then, with the information of $\lambda$ the third one estimates the optimal $k$, $\alpha_k$ and $\theta_k$ 
for the search with certainty buy using Eqs.~(\ref{eq:4}), (\ref{eq:5}) and (\ref{eq:16}). 
In the iterative search with certainty, by setting the parameter $\alpha_k$ on the circuit the third one
can find the marked state with certainty by $k$-time iterations. On the other hand, in the present effective
search,  by setting the parameters $k$, $\alpha_k$ and $\theta_k$ on the circuit the third one can find the marked
state by effective one oracle processing, where it can be possible to  devise the circuit so that 
the input parameters can be $\{k \ \ \rm{and}\ \ \alpha_k\}$ or 
$\{k \ \ \rm{and}\ \ \theta_k\}$ because $\alpha_k$ and $\theta_k$ are related to each other by Eq.~(\ref{eq:10}).  

Before ending this section, we consider a parallel processing system implied by the reduced form 
of Eq.~(\ref{eq:29}). It represents that the final search state 
$|\phi_k\rangle$ is given as a superposition of $|\phi_0\rangle$ and 
$|\phi_U\rangle \equiv U_N(-\alpha) |\phi_0\rangle$. 
This may imply a parallel processing scheme with two inputs ($|\phi_0\rangle$ and 
$|\phi_U\rangle$) and two outputs ($|\phi_k\rangle$ and an ancillary state $|\chi\rangle$).  
In  Appendix C we present an analysis of this scheme.  
We show that this two-channel processing can be decoupled into the single processing
of Fig.~1 that we examined in the present section and an ancillary single processing (see Eq.~(\ref{eq:C7})).

%\vspace{0.2in}
\section{Summary and discussion} 
\label{sec4}
We have derived an additive decomposition scheme of an iterative phase-matching  
search-algorithm of the Grover-type, in which 
a $k$ iterative search-process is, in the computational space, expressed as a  
linear combination of $k$ one-time searches followed by overall phase-rotations.
The phase rotation parameter $\theta$ is preliminarily determined by the fraction $\lambda$ 
of the targets and a matching phase $\alpha$.  
We emphasize that in the decomposition, the number of oracle operations remains the same. 
The decomposed form can be rewritten such that the final state is simply expressed as a 
superposition of the initial database state and a one-time searched state, where 
superposition coefficients consist of $\theta$ alone as shown in Eq.~(\ref{eq:26}).
We further showed that the decomposed form can be reduced to the form where  
the final state is expressed in terms of only {\it a single oracle operator} (without any diffusion operator; see Eq.~(\ref{eq:29})).
The decomposition holds for any $k$, $\alpha, \lambda$ and the number of qubit-registers $n$, 
as long as $\theta$ is determined by Eq.~(\ref{eq:10}).
Although it yields the norm of the final state correctly, the reduced operator $G^{I,k}_N(\alpha)$ itself
in the decomposed form of Eq.~(\ref{eq:26}) is not unitary in general.
Thus $G^{I,k}_N(\alpha)$ cannot directly be implemented as quantum 
mechanical evolution by unitary gates.
Therefore, by utilizing the advantage of $G^{I\!I,k}_N(\alpha)$ we proposed a unitary transformation that 
directly transforms the initial database state to the final state.
To determine the unitary transformation we only need the information on the one-time
oracle operation on the initial state~\cite{tdn14}.

%\newpage
For the exact search~\cite{toyama13} with the desired success probability $P_k(\lambda)=1$ for $0<\lambda \le 1$,
the construction of the unitary transformation can be much simplified because
the components of the final state vector have nonzero values only for the components corresponding 
to the target basis states.  
In their matrix representations, the unitary transformation is much simpler than the original $k$ iterative search operation.
As an example we illustrated this situation explicitly for the case of $n=3\  (N=8), k=2$ and one target. 
An effective reduction of the computational load is important in particular in the situation with a small $\lambda$
(namely, a small number of targets), in which the final state of the exact search algorithm becomes very simple.  
%\newpage
%\noindent 

Let us add that for the search problem where $\lambda$ is not given preliminarily, the multi-phase matching (MPM)
method \cite{toyama08, toyama09, yoder14, tan14} is useful. 
The MPM yields the success probability $P_k(\lambda)$ that is almost constant and unity over a 
wide range of $\lambda$, i.e., $P(\lambda) \simeq 1$ for $0< \lambda \leq 1$.
This MPM method enables searches with certainty with no information of $\lambda$.  
Instead, in MPM method the number of iterations $k$ is fixed preliminarily for an iterative search 
with an arbitrary $\lambda$, where $k$ is determined in finding the matched multi-phases. 
Therefore, for a large $\lambda$ the number of iterations $k$ of MPM is much larger than that of the exact search 
with $P_k(\lambda)=1$ reviewed in Sec. II. 
It would be meaningful to examine whether or not a similar scheme of reducing the computational 
load of the MPM method effectively is possible. 

\acknowledgments
This work was supported by the Japan Society for the Promotion of Sciences (JSPS: 16K05489) 
and Kyoto Sangyo University Research Grant.  
The authors wish to express their gratitude to Prof. Y. Nogami for his comments and suggestions.

\vspace{0.5in}

%\newpage
%\clearpage

\appendix

\section{AN ILLUSTRATION  OF THE  ADDITIVE DECOMPOSITION OF THE ITERATIVE 
GROVER SEARCH IN THE COMPUTATIONAL SPACE}
%\label{appen:a}
%\renewcommand{\theequation}{A\arabic{equation} }
%\setcounter{equation}{0}

We give an illustration of the additive decomposition scheme of the iterative Grover search. 
As an example, we consider the case of $n=3\ \ (N=8)$ and a target $|t_0\rangle=|000 \rangle$. 
In this case $\lambda=1/8$. Hence, the optimal number of searches for the exact search
is $k=2$ (see Eq.~(\ref{eq:5})). In the computational space, 
$|\phi_0\rangle = \frac{1}{\sqrt{8}} \left( 1, 1, 1, 1, 1, 1, 1, 1 \right)^{T}$ and
$|\omega_0\rangle = |000\rangle = \left( 1, 0, 0, 0, 0, 0, 0, 0 \right)^{T}$, where $T$ indicates {\it transposed}.
In the eight-dimensional computation space, the search operator $G_8(\alpha_2)$ is represented as
%----------------------------------eq(A1)-----------------------------------------------------------------------------------------------------------------
\begin{widetext}
\begin{eqnarray} \label{eq:A1}
\!\!\!\!\!\!\!\!\!G_8(\alpha_2)&=& W_8(-\alpha_2) U_8(\alpha_2)    \nonumber  \\
    &=&  \frac{1}{8} 
    \begin{pmatrix}
         e^{i\alpha_2}+7 & 1-e^{-i\alpha_2} & 1-e^{-i\alpha_2} & 1-e^{-i\alpha_2} & 1-e^{-i\alpha_2} & 1-e^{-i\alpha_2} & 1-e^{-i\alpha_2} & 1-e^{-i\alpha_2} \\
         e^{i\alpha_2}-1 & 1+7e^{-i\alpha_2} & 1-e^{-i\alpha_2} & 1-e^{-i\alpha_2} & 1-e^{-i\alpha_2} & 1-e^{-i\alpha_2} & 1-e^{-i\alpha_2} & 1-e^{-i\alpha_2} \\
         e^{i\alpha_2}-1 & 1-e^{-i\alpha_2} & 1+7e^{-i\alpha_2} & 1-e^{-i\alpha_2} & 1-e^{-i\alpha_2} & 1-e^{-i\alpha_2} & 1-e^{-i\alpha_2} & 1-e^{-i\alpha_2} \\
         e^{i\alpha_2}-1 & 1-e^{-i\alpha_2} & 1-e^{-i\alpha_2} & 1+7e^{-i\alpha_2} & 1-e^{-i\alpha_2} & 1-e^{-i\alpha_2} & 1-e^{-i\alpha_2} & 1-e^{-i\alpha_2} \\
         e^{i\alpha_2}-1 & 1-e^{-i\alpha_2} & 1-e^{-i\alpha_2} & 1-e^{-i\alpha_2} & 1+7e^{-i\alpha_2} & 1-e^{-i\alpha_2} & 1-e^{-i\alpha_2} & 1-e^{-i\alpha_2} \\
         e^{i\alpha_2}-1 & 1-e^{-i\alpha_2} & 1-e^{-i\alpha_2} & 1-e^{-i\alpha_2} & 1-e^{-i\alpha_2} & 1+7e^{-i\alpha_2} & 1-e^{-i\alpha_2} & 1-e^{-i\alpha_2} \\
         e^{i\alpha_2}-1 & 1-e^{-i\alpha_2} & 1-e^{-i\alpha_2} & 1-e^{-i\alpha_2} & 1-e^{-i\alpha_2} & 1-e^{-i\alpha_2} & 1+7e^{-i\alpha_2} & 1-e^{-i\alpha_2} \\
         e^{i\alpha_2}-1 & 1-e^{-i\alpha_2} & 1-e^{-i\alpha_2} & 1-e^{-i\alpha_2} & 1-e^{-i\alpha_2} & 1-e^{-i\alpha_2} & 1-e^{-i\alpha_2} & 1+7e^{-i\alpha_2} \\
       \end{pmatrix}.
\end{eqnarray}
\end{widetext}
%------------------------------------------------------------------------------------------------------------------------------------------------------------
With this $G_8(\alpha_2)$ of Eq.~(\ref{eq:A1}), the final state $|\phi_2\rangle = G^2_8(\alpha_2) |\phi_0\rangle$ is given as
%------------------------------------eq.(A2)--------------------------------
\begin{eqnarray} \label{eq:A2}
|\phi_2\rangle = \frac{\sqrt{2}}{64}
    \begin{pmatrix}
      - (3\cos\alpha_2+14)\cos\alpha_2 + 33 \\
       \ \ \ \ \ \ \ \ \ + i 4(\cos\alpha_2+7 )\sin\alpha_2 \\
      \cos^2\!\alpha_2+10\cos\alpha_2+5 \\
      \cos^2\!\alpha_2+10\cos\alpha_2+5 \\
      \cos^2\!\alpha_2+10\cos\alpha_2+5 \\
      \cos^2\!\alpha_2+10\cos\alpha_2+5 \\
      \cos^2\!\alpha_2+10\cos\alpha_2+5 \\
      \cos^2\!\alpha_2+10\cos\alpha_2+5 \\
      \cos^2\!\alpha_2+10\cos\alpha_2+5 \\
       \end{pmatrix}.
\end{eqnarray}
%-------------------------------------------------------------------------------------
The matching phase $\alpha_2$ for the exact search is given as (see Eq.~(\ref{eq:4})), 
%-------------------------------------eq.(A3)----------------------------------------
\begin{eqnarray} \label{eq:A3}
  \alpha_2  &=& \arccos\left[ 1- 8(1-\cos(\pi/5)) \right]  \nonumber  \\
    &=&\arccos(-5+2\sqrt{5}) .   
\end{eqnarray}
%------------------------------------------------------------------------------------------
With this  $\alpha_2$, Eq.~(\ref{eq:A2}) turns out to be
%-------------------------------------eq.(A4)-------------------------------------------
\begin{eqnarray}\label{eq:A4}
\!\!\!\!\!\!\!\!\!\!\!\! |\phi_2\rangle =  \frac{1}{\sqrt{8}} 
    \begin{pmatrix}
       2(\sqrt{5} -1) + i \sqrt{5\sqrt{5} - 11} \ \ (\sqrt{5} +1) \\
      0 \\
      0 \\
      0 \\
      0 \\
      0 \\
      0 \\
      0 \\
       \end{pmatrix}.
\end{eqnarray}
%------------------------------------------------------------------------------------------------
The modulus of the first component of Eq.~(\ref{eq:A4}) is unity, namely 
$| \frac{1}{\sqrt{8}} [2(\sqrt{5} -1) + i \sqrt{5\sqrt{5} - 11} \ \ (\sqrt{5} +1)] |=1$. 
Thus, an exact search is completed. 

Next, we consider the decomposed form, Eq.~(\ref{eq:21}),
%--------------------------------eq(A5)------------------------------------------------------------------------------------------
\begin{eqnarray}\label{eq:A5}
 |\phi_2\rangle &=& G_8^2(\alpha_2) |\phi_0\rangle  \nonumber \\
     &=&  \left[ e^{i\theta}G_8(\alpha_2) + e^{-i\theta}G_8(\alpha_2)+ e^{i \pi} I_8 \right] |\phi_0 \rangle. 
\end{eqnarray}
%----------------------------------------------------------------------------------------------------------------------------------- 
The first and second terms of the r.h.s. of Eq.~(\ref{eq:A5}) are
%----------------------------------eq.(A6)--------------------------------------------------------------------------------------
\begin{eqnarray}\label{eq:A6}
\!\!\!\!\!\!\!\!\!\!\!\!e^{\pm i \theta_2} G_{8}(\alpha_2) |\phi_0 \rangle = e^{\pm i \theta_2} \frac{\sqrt{2}}{32} 
    \begin{pmatrix}
       14+e^{i\alpha_2}-7e^{-i\alpha_2} \\  
       6+e^{i\alpha_2}+e^{-i\alpha_2} \\
       6+e^{i\alpha_2}+e^{-i\alpha_2} \\
       6+e^{i\alpha_2}+e^{-i\alpha_2} \\
       6+e^{i\alpha_2}+e^{-i\alpha_2} \\
       6+e^{i\alpha_2}+e^{-i\alpha_2} \\
       6+e^{i\alpha_2}+e^{-i\alpha_2} \\
       6+e^{i\alpha_2}+e^{-i\alpha_2} \\
       \end{pmatrix}.
\end{eqnarray}
%-------------------------------------------------------------------------------------------------------------------------------------------------------------
The third term in the r.h.s. of Eq.~(\ref{eq:A5}) is $e^{i \pi} I_8 |\phi_0 \rangle = -  \frac{1}{\sqrt{8}}(1, 1, 1, 1, 1, 1, 1, 1)^T$.
Accordingly, Eq.~(\ref{eq:A5}) is represented as
%----------------------------------eq.(A7)-----------------------------------------------------------------------------------------------------------------
\begin{eqnarray}
& &|\phi_2\rangle=\left[ e^{i\theta_2}G_8(\alpha_2) + e^{-i\theta_2}G_8(\alpha_2)+ e^{i \pi} I_8 \right] |\phi_0 \rangle   \nonumber \\
&=& \frac{\sqrt{2}}{8}
    \begin{pmatrix}
        \cos\theta_2 ( 7 - 3\cos\alpha_2 + i 4 \sin\alpha_2) - 2    \\  
       \cos\theta_2 (\cos\alpha_2 + 3)-2                      \\
       \cos\theta_2 (\cos\alpha_2 + 3)-2                       \\
       \cos\theta_2 (\cos\alpha_2 + 3)-2                      \\
       \cos\theta_2 (\cos\alpha_2 + 3)-2  \label{eq:A7} \\
       \cos\theta_2 (\cos\alpha_2 + 3)-2                      \\
       \cos\theta_2 (\cos\alpha_2 + 3)-2                       \\ 
       \cos\theta_2 (\cos\alpha_2 + 3)-2                       \\
       \end{pmatrix} .  
\end{eqnarray}
%--------------------------------------------------------------------------------------------------------------------------------------------------
From Eq.~(\ref{eq:10}), $\cos \theta_2 = (7 + \cos \alpha_2)/8$. By using this relation,  Eq.~(\ref{eq:A7}) is reduced to Eq.~(\ref{eq:A2}).
Furthermore, with the overall rotation phase $\theta_2 =\pi/ \left(2k +1 \right)=\pi/5$ and $\alpha_2$ of Eq.~(\ref{eq:A3}), Eq.~(\ref{eq:A7}) is reduced to Eq.~(\ref{eq:A4}). 
This illustration verifies that the additive decomposition formula of the state vector $|\phi_k \rangle$ of Eq.~(\ref{eq:26})
 is correct as an identity.

%\clearpage
%\vspace{1cm}

\section{A MATRIX REPRESENTATION OF THE UNITARY TRANSFORMATION $C_N[U_N(-\alpha_k), \theta_k]$  }

As an example we give a matrix representation of the unitary transformation 
$C_N[U_N(-\alpha_k), \theta_k]$ in the computational basis for the cace of the example
shown in Appendix A ($N=8 \ \ (n=3)$ and $k=2$). The matrix representation is compared with
that of the two-time iterative-search operator $G_N^2(\alpha_2)$.   
  
The transformation  $C_8[U_8(-\alpha_2), \theta_2]$ can be written as
%---------------------------------eq.(B1)---------------------------------------------------------------------
\begin{eqnarray}  \label{eq:B1}
&&C_8[U_8(-\alpha_2), \theta_2]  \nonumber \\
&&= |\phi^{(0)}_2 \rangle \langle \phi^{(0)}_0 | + |\phi^{(1)}_2 \rangle \langle \phi^{(1)}_0 | 
+ \cdots + |\phi^{(7)}_2 \rangle \langle \phi^{(7)}_0 |. 
\end{eqnarray}
%-----------------------------------------------------------------------------------------------------------------    
The matrix representation of Eq.~(\ref{eq:B1}) is obtained as follows.
The orthonormal set $|\phi^{(i)}_0\rangle \ \ (i=0,\dots, N-1)$ are given as 
%---------------------------------eq.(B2)(B3)(B4)(B5)----------------------------------------------------
\begin{eqnarray}  \label{eq:B2}
\!\!\!\!\!\!\!\!\!\!|\phi^{(0)}_0\rangle = |\phi_0\rangle= \frac{1}{\sqrt{8}} 
    \begin{pmatrix}
      1 \\
      1 \\
      1 \\
      1 \\
      1 \\
      1 \\
      1 \\
      1 \\
       \end{pmatrix},   \ \ \ \ \ \ \ \ \ \ \ \ \ \ \    
\!\!\!\!\!\!\!\!\!\!\!\!\!\!\!\!\!\!\!\!|\phi^{(1)}_0\rangle = \frac{1}{2\sqrt{14}}
    \begin{pmatrix}
          7 \\
         -1 \\
         -1 \\
         -1 \\
         -1 \\
         -1 \\
         -1 \\
         -1 \\
       \end{pmatrix},     
\end{eqnarray}
\begin{eqnarray}   \label{eq:B3}
\!\!\!\!\!\!\!\!\!\!|\phi^{(2)}_0\rangle = \frac{1}{\sqrt{42}}
    \begin{pmatrix}
         0 \\
      	 6 \\
        -1 \\
        -1 \\
        -1 \\
        -1 \\
        -1 \\
        -1 \\
       \end{pmatrix},   \ \ \    
       |\phi^{(3)}_0\rangle = \frac{1}{\sqrt{30}}
    \begin{pmatrix}
          0 \\
          0 \\
          5 \\
         -1 \\
         -1 \\
         -1 \\
         -1 \\
         -1 \\
       \end{pmatrix},        
\end{eqnarray}
\begin{eqnarray}      \label{eq:B4}
\!\!\!\!\!\!\!\!\!\!|\phi^{(4)}_0\rangle = \frac{1}{2\sqrt{5}}
    \begin{pmatrix}
         0 \\
         0 \\
         0 \\
         4 \\
        -1 \\
        -1 \\
        -1 \\
        -1 \\
       \end{pmatrix},   \ \ \ \ \ \ \   
       |\phi^{(5)}_0\rangle = \frac{1}{2\sqrt{3}}
    \begin{pmatrix}
          0 \\
          0 \\
          0 \\
          0 \\
          3 \\
         -1 \\
         -1 \\
         -1 \\
       \end{pmatrix},     
\end{eqnarray}
\begin{eqnarray}   \label{eq:B5}
\!\!\!\!\!\!\!\!\!\!|\phi^{(6)}_0\rangle = \frac{1}{\sqrt{6}}
    \begin{pmatrix}
        0 \\
        0 \\
        0 \\
        0 \\
        0 \\
        2 \\
       -1 \\
       -1 \\
       \end{pmatrix},   \ \ \ \ \ \ \    
       |\phi^{(7)}_0\rangle = \frac{1}{\sqrt{2}}
    \begin{pmatrix}
          0 \\
          0 \\
          0 \\
          0 \\
          0 \\
          0 \\
          1 \\
         -1 \\
       \end{pmatrix},  
\end{eqnarray}
%---------------------------------------------------------------------------------------------------------------------------------------------    
where orthonormal states $|\phi^{(i)}_0\rangle \ \ (i=1,\dots,7)$ were obtained by means of the 
Gram-Schmidt orthogonalization with subsidiarily prepared linearly independent seven vectors, 
$|\omega_0\rangle=(1, 0, 0, 0, 0, 0, 0, 0)^T, 
|\omega_1\rangle=(0, 1, 0, 0, 0, 0, 0, 0)^T, \ldots, |\omega_6\rangle=(0, 0, 0, 0, 0, 0, 1, 0)^T$.

The  $|\phi^{(0)}_2\rangle  \equiv |\phi_2\rangle$ is given by Eq.~(\ref{eq:A7}). 
On the other hand, we can also calculate it much more simply by the information on the oracle operation 
$U_8(-\alpha_2) |\phi_0 \rangle$ alone, by means of  $|\phi_2\rangle = \left[ g_2(\theta_2) I_N 
+ h_2(\theta_2) U_8(-\alpha_2) \right] |\phi_0\rangle$ of Eq.~(\ref{eq:29}).  
As shown in Eqs.~(\ref{eq:30})-(\ref{eq:32}), this formula yields 
%----------------------------------------------------eq.(B6)-------------------------------------------------------------------------------
\begin{eqnarray}    \label{eq:B6}
&&\!\!\!\!\!\!\!\!|\phi^{(0)}_2\rangle=|\phi_2\rangle  \nonumber \\  
&&\!\!\!\! =  \begin{pmatrix}  
          v_t  \\
          v_{nt}  \\
          v_{nt}  \\
          v_{nt} \\  
          v_{nt} \\
          v_{nt} \\
          v_{nt} \\
          v_{nt} \\         
       \end{pmatrix}     
         =  \frac{1}{\sqrt{8}}  
          \begin{pmatrix}  
          4\cos^2\theta_2 - 2\cos\theta_2 e^{-i\alpha_2} -1 \\
          4\cos^2\theta_2 - 2 \cos\theta_2 -1\\
          4\cos^2\theta_2 - 2 \cos\theta_2 -1 \\
          4\cos^2\theta_2 - 2 \cos\theta_2 -1 \\   
          4\cos^2\theta_2 - 2 \cos\theta_2 -1 \\
          4\cos^2\theta_2 - 2 \cos\theta_2 -1 \\
          4\cos^2\theta_2 - 2 \cos\theta_2 -1 \\
          4\cos^2\theta_2 - 2 \cos\theta_2 -1 \\         
       \end{pmatrix}.   
\end{eqnarray}
With the relation $\cos\theta_2=(7+\cos\alpha_2)/8$, this $|\phi_2\rangle$ can be reduced to Eq.~(\ref{eq:A7}).
Furthermore, with $\theta_2 = \pi/5$ and $\alpha_2$ of Eq.~(\ref{eq:A3}), Eq.~(\ref{eq:B6}) is reduced to Eq.~(\ref{eq:A4}).
Namely, $|\phi_2\rangle=(v_t, 0, 0, 0, 0, 0, 0, 0)^T = v_t |\omega_0\rangle$, where $v_t =  \frac{1}{\sqrt{8}} [4\cos^2\theta_2 - 2\cos\theta_2 e^{-i\alpha_2} -1] =  \frac{1}{\sqrt{8}}[ 2\sqrt{5}-2 + i \sqrt{-11+5 \sqrt{5}} \ \ (\sqrt{5}+1) ]$,  
of which modulus is unity.  Hence, the orthonormal set $|\phi^{(i)}_2\rangle \ \ (i=0,\dots,N-1)$ is given as
%----------eq.(B7),(B8),(B9)---------------------------------
\begin{eqnarray}    \label{eq:B7}
\!\!\!\!\!\!\!\!\!\!\! |\phi^{(0)}_2\rangle = |\phi_2\rangle =
    \begin{pmatrix}
          v_t \\
          0 \\
          0 \\
          0 \\
          0 \\
          0 \\
          0 \\
          0 \\
       \end{pmatrix}, 
|\phi^{(1)}_2\rangle = 
    \begin{pmatrix}
          0 \\
          1 \\
          0 \\
          0 \\
          0 \\
          0 \\
          0 \\
          0 \\
       \end{pmatrix}, 
 |\phi^{(2)}_2\rangle = 
    \begin{pmatrix}
          0 \\
      	  0 \\
         1 \\
         0 \\
         0 \\
         0 \\
         0 \\
         0 \\
       \end{pmatrix},  
\end{eqnarray}    
\begin{eqnarray}   \label{eq:B8}
\ \ \ |\phi^{(3)}_2\rangle = 
    \begin{pmatrix}
          0 \\
          0 \\
          0 \\
          1 \\
          0 \\
          0 \\
          0 \\
          0 \\
       \end{pmatrix},      
|\phi^{(4)}_2\rangle = 
    \begin{pmatrix}
          0 \\
          0 \\
          0 \\
          0 \\
          1 \\
          0 \\
          0 \\
          0 \\
       \end{pmatrix},  
|\phi^{(5)}_2\rangle = 
    \begin{pmatrix}
          0 \\
          0 \\
          0 \\
          0 \\
          0 \\
          1 \\
          0 \\
          0 \\
       \end{pmatrix},     
\end{eqnarray}    

 %\vspace{2cm}
 
\begin{eqnarray}    \label{eq:B9}
\!\!\!\!\!\!\!\!\!\!\!\!\!\!\!\!\!\!\!\!\!\!\!\!\!\!\!\!\!\!\!\!\!\!\!\!\!\!\!\!\!\!\!\!\!\!\!\!
\ \ \ \ \ \ \ \ |\phi^{(6)}_2\rangle = 
    \begin{pmatrix}
        0 \\
        0 \\
        0 \\
        0 \\
        0 \\
        0 \\
        1 \\
        0 \\
       \end{pmatrix},      
|\phi^{(7)}_2\rangle = 
    \begin{pmatrix}
          0 \\
          0 \\
          0 \\
          0 \\
          0 \\
          0 \\
          0 \\
          1 \\
       \end{pmatrix}.  
\end{eqnarray}
%-----------------------------------------------------------------------------------------------------------------------------------------  
The completeness conditions $\sum_{i=0}^{7} | \phi^{(i)}_0 \rangle \langle \phi^{(i)}_0 | = I_8$ 
and $\sum_{i=0}^{7} | \phi^{(i)}_2 \rangle \langle \phi^{(i)}_2 | = I_8$  are easily verified. This guarantees 
the unitarity of $C_8[U_8(-\alpha_2), \theta_2]$.
With these orthonormal sets of Eqs.~(\ref{eq:B2})-(\ref{eq:B5}) and 
Eqs.~(\ref{eq:B7})-(\ref{eq:B9}), 
the unitary transformation $C_8[U_8(-\alpha_2), \theta_2]$  is given as,
%-----------------------------eq.(B10)---------------------------------------------------------------------------------------------------
\begin{widetext}
 \vspace{1cm}
\begin{eqnarray}  \label{eq:B10}
&&C_8[U_8(-\alpha_2), \theta_2]  \nonumber    \\
&&= v_t |\omega_0 \rangle \langle \phi_0 | + |\omega_1 \rangle \langle \phi^{(1)}_0 | 
       + \cdots + |\omega_7 \rangle \langle \phi^{(7)}_0 |      \\ \nonumber
&&=    
    \begin{pmatrix}
     v_t/(\sqrt{8}) & v_t/(\sqrt{8}) & v_t/(\sqrt{8}) & v_t/(\sqrt{8}) & v_t/(\sqrt{8}) & v_t/(\sqrt{8}) & v_t/(\sqrt{8}) 
     & v_t/(\sqrt{8}) \\
     \sqrt{7}/(\sqrt{8})    &  -1/(2\sqrt{14}) & -1/(2\sqrt{14}) & -1/(2\sqrt{14}) & -1/(2\sqrt{14}) & -1/(2\sqrt{14}) 
     & -1/(2\sqrt{14}) & -1/(2\sqrt{14}) \\
     0 & \sqrt{6}/\sqrt{7} & -1/\sqrt{42} & -1/\sqrt{42} & -1/\sqrt{42} & -1/\sqrt{42} & -1/\sqrt{42} & -1/\sqrt{42}  \\
     0 & 0 & \sqrt{5}/\sqrt{6} & -1/\sqrt{30} & -1/\sqrt{30} & -1/\sqrt{30} & -1/\sqrt{30} & -1/\sqrt{30}  \\
     0 & 0 & 0 & 2/\sqrt{5} & -1/(2\sqrt{5}) & -1/(2\sqrt{5}) & -1/(2\sqrt{5}) & -1/(2\sqrt{5}) \\
     0 & 0 & 0 & 0 & \sqrt{3}/2 & -1/(2\sqrt{3}) & -1/(2\sqrt{3}) & -1/(2\sqrt{3})  \\
     0 & 0 & 0 & 0 & 0 & \sqrt{2}/\sqrt{3} & -1/\sqrt{6} & -1/\sqrt{6} \\
     0 & 0 & 0 & 0 & 0 & 0 & 1/\sqrt{2} & -1/\sqrt{2}  \\
    \end{pmatrix}.
\end{eqnarray} 
\end{widetext}
%--------------------------------------------------------------------------------------------------------------------------------------    
It can easily be confirmed that $C^{\dag}_8 C_8 = C_8 C^{\dag}_8= I_8$ holds. 
Also, $|\phi_2\rangle =  C_8 |\phi_0\rangle$ is obvious because the sum of the elements of the first row of 
Eq.~(\ref{eq:B9}) is $ \sqrt{8} v_t$ and the sum of the elements of each row other than the first row is zero. 
Note that  $C_8[U_8(-\alpha_2), \theta_2]$ is constructed with the information of the single oracle operation 
$U_8(-\alpha_2) |\phi_0\rangle$ only. The diffusion operator $W_8(\alpha_2)$ is not needed in constructing 
$C_8[U_8(-\alpha_2), \theta_2]$.

On the other hand, the iterative search operator $G^2_8(\alpha_2)$ obtained from Eq.~(\ref{eq:A1}) 
is much more complicated than that of $C_8[U_8(-\alpha_2), \theta_2]$ as shown below
%-----------------------------eq.(B11)---------------------------------------------------------------------------------------------------
%\begin{widetext}
\begin{eqnarray}  \label{eq:B11}
&&G^2_8 \left( \alpha_2 \right)  =    
    \begin{pmatrix}
     a  &  a  &  a  &  a  &  a  &  a  &  a  &  a    \\
     b  &  c  &  d  &  d  &  d  &  d  &  d  &  d    \\
     b  &  d  &  c  &  d  &  d  &  d  &  d  &  d    \\
     b  &  d  &  d  &  c   &  d  &  d  &  d  &  d   \\
     b  &  d  &  d  &  d  &  c   &  d  &  d  &  d   \\
     b  &  d  &  d  &  d  &  d  &  c   &  d  &  d   \\
     b  &  d  &  d  &  d  &  d  &  d  &   c  &  d   \\
     b  &  d  &  d  &  d  &  d  &  d  &   d  &  c   \\
    \end{pmatrix},
\end{eqnarray}
%\end{widetext}
%\newpage
where 
%\newpage
%-----------------------------eq.(B12)(B13)(B14)(B15)-----------------------------------------------------------------------------------------
\begin{eqnarray}  
&&\!\!\!\!\!\! a =  \frac{v_t}{ \sqrt{8}},    \label{eq:B12}  \\
&&\!\!\!\!\!\! b =  \frac{1}{8} \left[ 2 - 2 \sqrt{5}  + i \sqrt{-11+5\sqrt{5}} \left( 1 + \sqrt{5} \right)  \right],    \label{eq:B13} \\
&&\!\!\!\!\!\! c =  \frac{1}{8} \left[ 610 - 274\sqrt{5}  \right.    \nonumber  \\
&&  \ \ \ \ \ \ \ \ \ \ \ \ \ \ \ \ \                      \left.  + i \sqrt{-11+ 5\sqrt{5}} \left( 137 - 55 \sqrt{5} \right) \right],    \label{eq:B14}  \\
&&\!\!\!\!\!\! d =  \frac{1}{8} \left[ -102 + 46\sqrt{5}   \right.  \nonumber  \\
&&  \ \ \ \ \ \ \ \ \ \ \ \ \ \ \ \ \ \ \ \ \       \left.  - i \sqrt{-11+ 5\sqrt{5}} \left( 23 - 9 \sqrt{5} \right) \right].    \label{eq:B15}
\end{eqnarray}
%------------------------------------------------------------------------------------------------------------------------------------------------------
%\newline
\noindent In the this representation,  the first row of $G^2_8(\alpha_2)$ is the same as that of $C_8[U_8(-\alpha_2), \theta_2]$. 
Also, the sum of the elements in each row except the first row is zero. 
Hence, obviously $G^2_8(\alpha_2) |\phi_0\rangle = C_8[U_8(-\alpha_2), \theta_2] |\phi_0\rangle$ holds for 
$|\phi_0\rangle = \frac{1}{\sqrt{8}} \left( 1, 1, 1, 1, 1, 1, 1, 1 \right)^{T}$.  \\

%\newpage
%\newline
%\clearpage

\section{A PARALLEL PROCESSING SCHEME SUGGESTED BY THE REDUCED FORM OF Eq.~(\ref{eq:29})}
The reduced form of Eq.~(\ref{eq:29}) might imply the parallel processing scheme shown in Fig.~2. 
Mathematically it is understood that the input state is the tensor 
product state of $|\phi_0\rangle$ and $|\phi_U\rangle$, where each state is $N$-dimensional.
The input state is thus $N^2$-dimensional. The output state is expressed as a tensor product state of the 
final-search state $|\phi_k\rangle$ and an ancillary $N$-dimensional state $|\chi \rangle$, which is needed for 
consistency of dimensions between the input and output states. 
The state $|\chi \rangle$ can be taken appropriately
in constructing unitary transformation. 
In the scheme it is assumed that the superposition coefficients $g_k(\theta)$ and $h_k(\theta)$ are incorporated
in the transformation $C_N^P$.  
Hence the parallel processing scheme shown in Fig.~2 can be mathematically expressed as 
%------------------------------eq.(C1)--------------------------------------------------------------------------------------------------------
\begin{equation}  
   |\phi_2\rangle \otimes |\chi\rangle 
   = C_N^P[U_N(-\alpha_k), \theta_k] \left( |\phi_0\rangle \otimes  |\phi_U \rangle \right).  \label{eq:C1} 
\end{equation}
%----------------------------------------------------------------------------------------------------------------------------------------------
We can write the unitary transformation $C_N^P$ as
%------------------------------eq.(C2)-------------------------------------------------------------------------------------------------------
\begin{eqnarray} 
C_N^P[U_N(-\alpha_k), \theta_k] = \sum_{j=0}^{N^2-1}  |\Psi_2^{(j)} \rangle \langle \Psi_0^{(j)} |,  \label{eq:C2}  
\end{eqnarray}
%----------------------------------------------------------------------------------------------------------------------------------------------- 
where $|\Psi_0^{(j)} \rangle$ and $| \Psi_2^{(j)} \rangle$ are respectively $N^2$-dimensional orthonormal sets,
%They are respectively taken to be orthonormal, i.e.,
namely $\langle \Psi_0^{(i)} | \Psi_0^{(j)} \rangle = \delta_{ij}$ and $\langle \Psi_2^{(i)} | \Psi_2^{(j)} \rangle = \delta_{ij}$.
Thus $C_N^P$ is obviously unitary because the completeness holds for each set $ |\Psi_2^{(j)} \rangle ( j=0, \ldots, N^2-1)$ 
and  $ |\Psi_0^{(j)} \rangle ( j=0, \ldots N^2-1)$.  The $| \Psi_0^{(j)} \rangle$ and $| \Psi_2^{(j)} \rangle$ can respectively be
tensor product states such as  
%------------------------------eq.(C3)(C4)-----------------------------------------------------------------------------------------
\begin{eqnarray} 
&& | \Psi_0^{(j)} \rangle = |\phi_0^{(p)}\rangle \otimes |\phi_U^{(q)}\rangle,   \label{eq:C3}  \\
&& | \Psi_2^{(j)} \rangle = |\phi_2^{(p)}\rangle \otimes |\chi^{(q)}\rangle,   \label{eq:C4}  
\end{eqnarray}
%---------------------------------------------------------------------------------------------------------------------------------- 
where $(p, q)=0, \ldots, N-1$ and $|\phi_0^{(0)}\rangle \equiv |\phi_0\rangle$, $|\phi_2^{(0)}\rangle \equiv |\phi_2\rangle$, 
$|\phi_U^{(0)} \rangle \equiv |\phi_U\rangle$ and $|\chi^{(0)}\rangle \equiv |\chi \rangle$. 
Further, $|\phi_2^{(p)}\rangle$, $|\phi_U^{(q)}\rangle$ and $|\chi^{(q)}\rangle$
are respectively taken to be orthonormal to each other, i.e., $\langle \phi_0^{(p)} | \phi_0^{(q)}\rangle=\delta_{pq}$, 
$\langle \phi_2^{(p)} | \phi_2^{(q)}\rangle=\delta_{pq}$, $\langle \phi_U^{(p)} | \phi_U^{(q)}\rangle=\delta_{pq}$ and 
$\langle \chi^{(p)} | \chi^{(q)}\rangle=\delta_{pq}$.    
By these definitions, the orthonormality $\langle \Psi_0^{(i)} | \Psi_0^{(j)} \rangle = \delta_{ij}$ 
and $\langle \Psi_2^{(i)} | \Psi_2^{(j)} \rangle = \delta_{ij}$ are guaranteed. 
%------------------------------------------------ fig2 -----------------------------------------------------------------------------------------
\begin{widetext}
\begin{figure}[ht]
\begin{minipage}{2.\linewidth}
  \centering
   \resizebox{5.in}{!}{\includegraphics{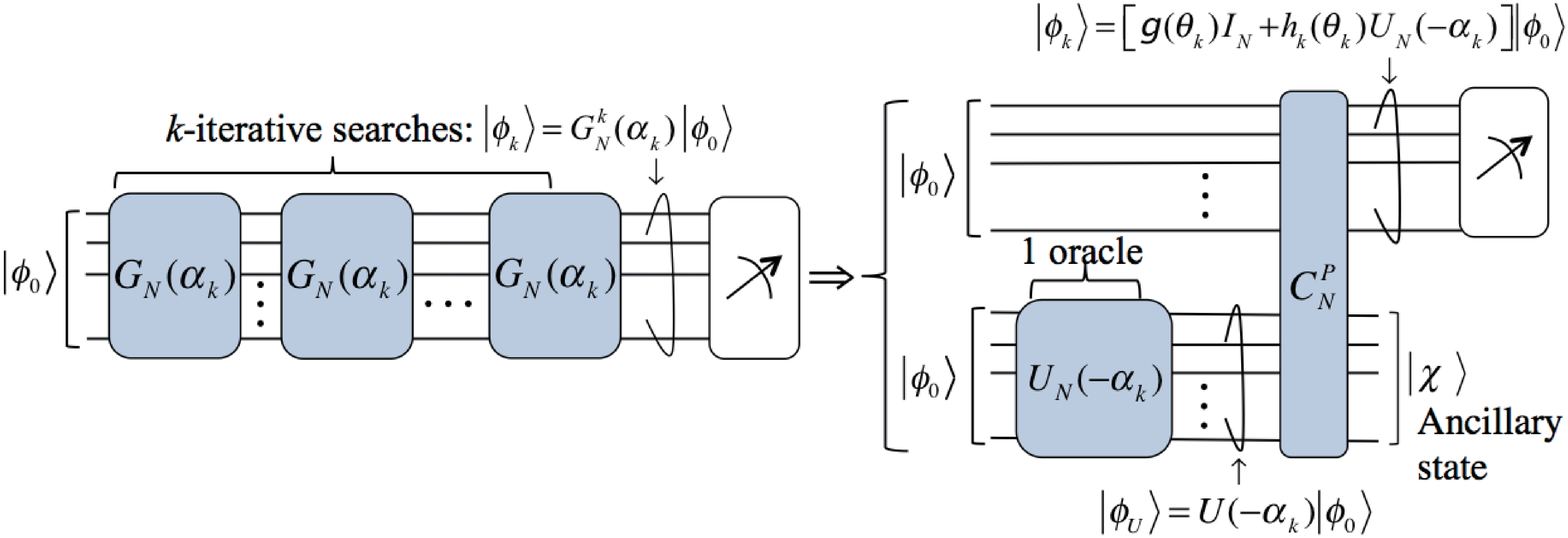}}
   \caption{ (Color online)  
   Parallel processing scheme implied by the reduced form of Eq.~(\ref{eq:29}), i.e.,
   $|\phi_k\rangle = \left[ g_k(\theta) I_N + h_k(\theta) U_N(-\alpha) \right] |\phi_0 \rangle 
   \equiv g_k(\theta) |\phi_0\rangle + h_k(\theta) |\phi_U\rangle$, where
   $|\phi_U\rangle \equiv U_N(-\alpha) |\phi_0 \rangle $.  
   The ancillary state $|\chi\rangle$ is needed for consistency of the dimensions between the input and output states, 
   which we can take appropriately in constructing a unitary transformation $C_N^P[U_N(-\alpha_k), \theta_k]$. 
   }
 \label{fig:2}
 \end{minipage}
\end{figure}
\end{widetext}
%---------------------------------------------------------------------------------------------------------------------------------------------

In the following we discuss explicitly the structure of the unitary transformation $C_N^P$ of Eq.~(\ref{eq:C2}). 
For simplicity we examine $C_N^P$ for the case of $N=2 \ \ (n=1)$.  The generalization to any $N$ is straightforward. 
The $C_N^P$ can be written as,  \
%------------------------------eq.(C5)-----------------------------------------------------------------------------------------
\begin{eqnarray} 
&C_N^P[U_N(-\alpha_k), \theta_k] &
       =  \sum_{j=0}^{2^2-1}  |\Psi_2^{(j)} \rangle \langle \Psi_0^{(j)} |    \nonumber  \\
       &=&\!\!\!\!\!\!\!\!\!\!\!\!\!\!\!\!\!\! \left( |\phi_2^{(0)}\rangle \otimes |\chi^{(0)} \rangle \right)
              \left(  \langle\phi_0^{(0)}| \otimes  \langle \phi_U^{(0)}|     \right)        \nonumber  \\
       &+&\!\!\!\!\!\!\!\!\!\!\!\!\!\!\!\!\!\! \left(  |\phi_2^{(0)}\rangle \otimes |\chi^{(1)}\rangle  \right)
              \left( \langle\phi_0^{(0)}| \otimes  \langle \phi_U^{(1)}|      \right)     \label{eq:C5}  \nonumber  \\           
       &+&\!\!\!\!\!\!\!\!\!\!\!\!\!\!\!\!\!\! \left( |\phi_2^{(1)}\rangle \otimes |\chi^{(0)}\rangle    \right)
               \left( \langle\phi_0^{(1)}| \otimes  \langle \phi_U^{(0)}|   \right)    \nonumber  \\
       &+&\!\!\!\!\!\!\!\!\!\!\!\!\!\!\!\!\!\!  \left( |\phi_2^{(1)}\rangle \otimes |\chi^{(1)}\rangle   \right)
               \left( \langle\phi_0^{(1)}| \otimes  \langle \phi_U^{(1)}|   \right).                  
\end{eqnarray}
%---------------------------------------------------------------------------------------------------------------------------------------
Equation (\ref{eq:C5}) is manipulated as
%------------------------------eq.(C6)(C7)-----------------------------------------------------------------------------------------
\begin{eqnarray}
\!\!\! C_N^P[U_N(-\alpha_k), \theta_k] &=& |\phi_2^{(0)}\rangle \langle\phi_0^{(0)}|   \otimes   |\chi^{(0)}\rangle \langle \phi_U^{(0)}|    \nonumber  \\
       &+& |\phi_2^{(0)}\rangle \langle\phi_0^{(0)}|   \otimes   |\chi^{(1)}\rangle \langle \phi_U^{(1)}|     \label{eq:C6}    \\
       &+& |\phi_2^{(1)}\rangle \langle\phi_0^{(1)}|   \otimes   |\chi^{(0)}\rangle \langle \phi_U^{(0)}|     \nonumber  \\ 
       &+& |\phi_2^{(1)}\rangle \langle\phi_0^{(1)}|   \otimes   |\chi^{(1)}\rangle \langle \phi_U^{(1)}|     \nonumber   
%       &=& \left( |\phi_2^{(0)}\rangle \langle\phi_0^{(0)}| + |\phi_2^{(1)}\rangle \langle\phi_0^{(1)}| \right)   \nonumber \\
%  &\otimes& \left( |\chi^{(0)}\rangle \langle\phi_U^{(0)}| + |\chi^{(1)}\rangle \langle \phi_U^{(1)}| \right).   \label{eq:C7}
\end{eqnarray}
%--------------------------------------------------------------------------------------------------------------------------------------------
%------------------------------continued eq.(C7)-----------------------------------------------------------------------------------------
\begin{eqnarray}
\ \ \ \ \ \ \ \ \ \ \ \ \ \ \ \ \ \ \ &=& \left( |\phi_2^{(0)}\rangle \langle\phi_0^{(0)}| + |\phi_2^{(1)}\rangle \langle\phi_0^{(1)}| \right)   \nonumber \\
  &\otimes& \left( |\chi^{(0)}\rangle \langle\phi_U^{(0)}| + |\chi^{(1)}\rangle \langle \phi_U^{(1)}| \right).   \label{eq:C7}
\end{eqnarray}
%--------------------------------------------------------------------------------------------------------------------------------------------
The final form of Eq.~(\ref{eq:C7}) implies that the states $|\phi_0\rangle\rangle$ and $|\phi_U\rangle$ in the input 
channel are separately transformed to the output states  $|\phi_2\rangle$ and $|\chi \rangle$,  namely 
%------------------------------eq.(C8)----------------------------------------------------------------------------------------------------
\begin{eqnarray}
&&\!\!\!\!\!\!\!\!\!\!\!\!C_N^P[U_N(-\alpha_k), \theta_k] \left( |\phi_0^{(0)}\rangle \otimes |\phi_U^{(0)}\rangle \right)  \nonumber \\
&=& \left[ \left( |\phi_2^{(0)}\rangle \langle\phi_0^{(0)}| + |\phi_2^{(1)}\rangle \langle\phi_0^{(1)}| \right) \right.     \nonumber \\
&\ \ \ \ \ \ \otimes& \left. \left( |\chi^{(0)}\rangle \langle\phi_U^{(0)}| + |\chi^{(1)}\rangle \langle \phi_U^{(1)}| \right) \right]
  \left( |\phi_0^{(0)}\rangle \otimes |\phi_U^{(0)}\rangle \right)    \nonumber  \\
&=&   |\phi_2^{(0)}\rangle \otimes |\chi^{(0)}\rangle.   \label{eq:C8}
\end{eqnarray}  
%--------------------------------------------------------------------------------------------------------------------------------------------
\\
This means that the two-channel parallel-processing scheme suggested by the reduced form of 
Eq.~(\ref{eq:29}) is equivalent to the decoupled two one-channel schemes for the inputs $|\phi_0\rangle$ and  
$|\phi_U\rangle$.
This situation of the decoupling of two channels is the same even if the channels 1 and 2 in the initial state are exchanged.

%\pagebreak
%\clearpage


\begin{thebibliography}{33}
\bibitem{tdn13}{ The $\theta_k$ of Eq.~(\ref{eq:16}) is the same as Eq.~(2.10) of Ref.~\cite{toyama13}
for the exact search.}
\bibitem{tdn14} {Grover discusses quantum searches by a single query~\cite{grover97a}.  He indicates that such more elaborate queries require a number of preprocessing and postprocessing steps that is greater than $O(N\log N)$.}
\end{thebibliography}
\end{document}